\documentclass[aps,pre,reprint]{revtex4-2} 

\usepackage{tikz}
\usepackage{amsmath,amssymb} 
\usepackage{graphicx} 
\usepackage{hyperref} 

\usepackage{txfonts}
\usepackage{bm,braket}

\newcommand{\p}{\partial}
\newcommand{\bx}{\bm{\mathrm{X}}}
\newcommand{\bmu}{\bm{\mu}}
\newcommand{\bW}{\bm{\mathrm{W}}}
\newcommand{\bn}{\bm{n}}
\newcommand{\bsigma}{\bm{\sigma}}
\newcommand{\diag}{\mathrm{diag}}

\newcommand{\diff}{\mathrm{d}}

\begin{document}

\title{Permutation of Tensor-Train Cores for Computing Moments on Stochastic Differential Equations}

\author{Kayo Kinjo$^1$,
 Rihito Sakurai$^2$,
 Tatsuya Kishimoto$^1$,
 and Jun Ohkubo$^1$}
\affiliation{$^1$Graduate School of Science and Engineering,
Saitama University,
Sakura, Saitama 338–8570,
Japan\\
$^2$ Department of Physics, The University of Tokyo, Bunkyo, Tokyo 113-0033, Japan}


\begin{abstract}
    Tensor networks, particularly the tensor train (TT) format, have emerged as powerful tools for high-dimensional computations in physics and computer science. In solving coupled differential equations, such as those arising from stochastic differential equations (SDEs) via duality relations, ordering the TT cores significantly influences numerical accuracy.
In this study, we first systematically investigate how different orderings of the TT cores affect the accuracy of computed moments using the duality relation in stochastic processes. 
Through numerical experiments on a two-body interaction model, we demonstrate that specific orderings of the TT cores yield lower relative errors, particularly when they align with the underlying interaction structure of the system. Motivated by these findings, we then propose a novel quantitative measure, \textit {score}, which is defined based on an ordering of the TT cores and an SDE parameter set. While the score is independent of the accuracy of moments to compute by definition, we assess its effectiveness by evaluating the accuracy of computed moments. Our results indicate that orderings that minimize the score tend to yield higher accuracy.
 This study provides insights into optimizing orderings of the TT cores, which is essential for efficient and reliable high-dimensional simulations of stochastic processes.
\end{abstract}

\maketitle

\section{Introduction} \label{sec:intro}
\begin{figure}[bp]
    \centering
    \begin{tikzpicture}
        \foreach \i [count=\j from 1] in {1,2,3,4} {
            \node[circle, draw=green!40!black, thick, minimum size=1cm] 
            (U\i) at ({1.5*\j}, 2.5) {$\mathcal{T}^{(\i)}$};
        }

        \foreach \i [count=\j from 1] in {1,2,3,4} {
            \node[circle, draw=red!40!black, thick, minimum size=1cm]
            (W\i) at ({1.5*\j}, 0) {$\mathcal{T}^{(\i)}$};
        }
        
        \foreach \i/\j [count=\k from 1] in {1/2,2/3,3/4} {
            \draw[red, thick] (W\i) -- (W\j) 
            node[midway, above] {$r_{\k}$};
        }

        \foreach \i/\j [count=\k from 1] in {1/2,2/3,3/4} {
            \draw[blue, thick] (U\i) -- (U\j) 
            node[midway, above] {$r_{\k}$};
        }

        \foreach \i [count=\j from 1] in {1,2,3,4} {
            \draw[black] (U\i) --++ (0,-0.8) node[right] {$i_{\j}$};
        }

        \foreach \i [count=\j from 1] in {1,2,3,4} {
            \draw[black] (W\i) --++ (0,+0.8) node[right] {$n_{\j}$};
            \draw[black] (W\i) --++ (0,-0.8) node[right] {$m_{\j}$};
        }
    \end{tikzpicture}
    \caption{(Color online) Illustration of the tensor-train (TT) format. The top row represents the TT as expressed in Eq.~\eqref{eq:def_mps} for $d=4$, while the bottom row represents the TT operator (TTO) as expressed in Eq.~\eqref{eq:def_mpo} for $d=4$, with independent physical indices. The circles represent the TT cores.}
    \label{fig:MPO_MPS}
\end{figure}
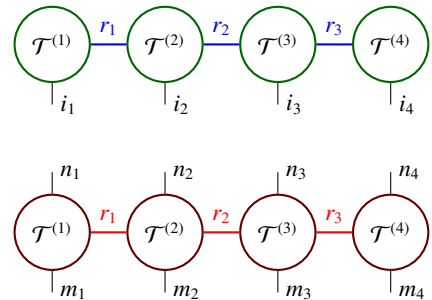

Tensor networks are widely used especially for physics and computer science. The tensor train (TT) and the tensor train operator (TTO), illustrated in Fig.~\ref{fig:MPO_MPS}, are one of the simplest and most practical tensor network representations. While the TT format as a formal tensor decomposition was introduced in the 2000s \cite{oseledets2009new,oseledets2011tensor}, its conceptual foundation dates back several decades. 
The fundamental idea underlying the TT format appeared implicitly in the 1970s on constructing the exact energy eigenstates of a quantum spin chain model \cite{faddeev1979quantum}, a method now known as the algebraic Bethe ansatz \cite{korepin1997quantum}. Subsequently, Murg, Korepin and Verstraete constructed the Bethe eigenstates with the MPS \cite{murg2012algebraic}. They showed that when they obtained Bethe eigenstates by applying a product of creation operators to a vacuum state, the orderings of the creation operators affected the entanglement entropy of the MPS, despite the fact that the creation operators commute theoretically, which means that the orderings should be arbitrary.
In the 1990s, the matrix product ansatz was introduced in some exactly solvable models, such as a stochastic process \cite{derrida1993exact} and a quantum spin chain model \cite{klumper1993matrix}. In physics, matrix product states (MPSs) and matrix product operators (MPOs) have been widely used to describe the tensor structure shown in Fig.~\ref{fig:MPO_MPS}.


Beyond exactly solvable models in physics, the TT format has emerged as a powerful computational tool for efficiently approximating ground states in numerical simulations of quantum many-body systems, most notably through the density matrix renormalization group (DMRG) \cite{white1992dmrg}. 
As the state space grows exponentially with the number of particles, the compression of the quantum states becomes essential for feasible numerical simulations.
While originally developed for quantum systems, the TT format provides a general framework applicable across various fields in applied mathematics and computational science. Although MPS and MPO are more common in physics, we use TT and TTO throughout this paper for consistency and generality.

The TT format has since been applied to diverse fields, including image processing, neural networks, and solving coupled differential equations \cite{zhao2016tensor, sugishita2024extraction, gao2020compressing,holtz_als,Dolgov2012TensorproductAT,dolgov2019tensor,richter2021solving}.
In particular, Richter, Sallandt and N{\"u}sken employed the TT format to solve backward stochastic differential equations (SDEs) derived from corresponding parabolic partial differential equations (PDEs), where the relationship between these SDEs and PDEs is established through duality relations \cite{richter2021solving}.

Duality relations between stochastic processes also have been used to compute the moments of SDEs without Monte Carlo sampling. When the initial distribution of the Fokker-Planck equation is given by a Dirac delta function $\delta(\bm x-\bm x_0)$, the moments associated with the stochastic process at time $t'$ is equivalent to the solution of the backward Kolmogorov equation evaluated at $\bm x= \bm x_0$ and $t=t'$.
This property enables efficient computation of moments for arbitrary initial values $\bm x_0$, making it particularly advantageous for data assimilation applications \cite{ohkubo_kalman}. 
Furthermore, since the dual process can be systematically derived from the governing SDEs, this approach provides a direct method for computing statistical moments without relying on Monte Carlo sampling.

However, applying this method to high-dimensional systems faces a major computational challenge: the curse of dimensionality. 
The cost of solving the backward Kolmogorov equation grows exponentially with the number of dimensions, making conventional approaches intractable. To address this issue, combinatorial techniques have been introduced to develop efficient algorithms that mitigate the computational burden of high-dimensional problems \cite{ohkubo2021combinatorics, ohkubo2023statistics}.

        



In this research, we achieve the following three contributions toward computing moments of SDEs using the TT format. Our first contribution is the application of the TT format to solve the backward Kolmogorov equation for the first time, which, to the best of our knowledge, is the first attempt in the context of SDEs. We also evaluate two of the most well-known algorithms for solving the backward Kolmogorov equation in this framework. Using the TT format enables us to compute moments of stochastic processes for systems up to 50 dimensions ($d=50$). In contrast, the conventional Crank-Nicolson (CN) method, without TT format, is limited to only 4 dimensions ($d=4$) under the same computational resources. 
For solving high-dimensional differential equations in the TT format, the alternating linear scheme (ALS) algorithm \cite{holtz_als} is widely used. This method transforms the original system into a sequence of smaller-dimensional linear equations, where the solution of each equation corresponds to each TT core (illustrate as circles in Fig.~\ref{fig:MPO_MPS}) of the solution. When applied to the backward Kolmogorov equation, this reduction enables efficient computation by exploiting the low-rank structure of the solution.
Inspired by the DMRG \cite{white1992dmrg} in physics, the modified ALS (MALS) algorithm further optimizes TT ranks (illustrate as horizontal edges in Fig.~\ref{fig:MPO_MPS}) by updating two adjacent TT cores simultaneously. However, our findings demonstrate that ALS alone is sufficient for accurately computing the moments of SDEs.
%


Our second contribution explores how the ordering of TT cores affects the accuracy of computed moments. This has not yet been investigated in the context of stochastic systems. This is because previous studies have primarily focused on systems with relatively simple interactions, such as nearest-neighbor interactions to solve high-dimensional differential equations \cite{dolgov2015simultaneous, gelss2017nearest}. In such cases, the optimal ordering of the TT cores is straightforward: TT cores are arranged so that interacting variables are adjacent, meaning the $n$-th TT core naturally corresponds to the $n$-th variable.
However, coupled differential equations derived from SDEs typically exhibit more complex interactions that extend beyond nearest-neighbor assumptions.  In these cases, it is not immediately clear how the orderings of the TT cores affect the accuracy of computed moments. To address this, we systematically analyze the relationship between ordering of the TT cores and numerical accuracy by comparing computed moments against Monte Carlo sampling results. Our findings demonstrate that ordering of the TT cores significantly influences accuracy, emphasizing the need for an ordering strategy that accounts for interaction complexity.
While some studies have explored tensor network structures that reflect underlying interaction topologies \cite{chen2024tnrecovery, murg2015tree, li2022permutation}, these approaches focus on determining the optimal ordering of the TT cores for a given tensor. In contrast, our research investigates the effect of ordering of the TT cores on solutions expressed in the TT format obtained by the ALS.

The final contribution of this work is the introduction of the \textit{score}, a novel metric designed to systematically indicate the optimal ordering of the TT cores. Specifying the ordering of the TT cores that yields the most accurate computed moments is a daunting task, particularly in high-dimensional systems.  Hence, based on the insights gained from our numerical experiments on how an ordering of the TT cores affects the accuracy of the computed moments, we propose the score as a guiding metric for improving the performance on computing the moments of SDEs.

The remainder of this paper is structured as follows. In Sect.~\ref{sec:method}, we introduce the methodology for computing statistical moments without Monte Carlo sampling by formulating the backward Kolmogorov equation. In Sect.~\ref{sec:time_evolution}, we present the application of the TT format to solving the backward Kolmogorov equation using the ALS algorithm, followed by a comparison of two numerical methods: ALS and MALS. In Sect.~\ref{sec:order_results}, we analyze how orderings of the TT cores influences the accuracy of computed moments, using the noisy Lotka-Volterra model as an example. Finally, in Sect.~\ref{sec:score}, we propose a novel \textit{score} metric to quantify orderings of the TT cores, defined by an ordering and a parameter set of an SDE, and assess its effectiveness based on the accuracy of computed moments.

\section{Computation of moments without Monte Carlo sampling} \label{sec:method}
We consider a stochastic differential equation (SDE) of $\bx (t) \in \mathbb{R}^d$: 
    \begin{align}
        \diff\bx = \bmu (\bm x) \diff t + \bsigma(\bm x) \diff \bW(t),
        \label{eq:sde_highd}
    \end{align}
    where $\bW(t)$ is a Wiener process, $\bmu(\bm x)$ is a vector of drift coefficient functions, and $\bsigma(\bm x)$ is a matrix of diffusion coefficient functions. Although it is common to employ Monte Carlo sampling to evaluate a moment of the process, $\mathbb{E}_{\bx(t)}\left[\bx^{\bn}\right]:=\mathbb{E}_{\bx(t)}\left[\mathrm{X}_1^{n_1}\mathrm{X}_2^{n_2}\cdots \mathrm{X}_d^{n_d}\right]$, one can compute it in different deterministic manners.
    Specifically, the duality relation in the stochastic process \cite{ohkubo2019duality} reduces the computation of the moments to solving coupled differential equations. 
    %
    We will derive the coupled diffrential equations to solve in this subsection, by starting with
    the Fokker-Plank equation of the SDE in Eq.~\eqref{eq:sde_highd},
    \begin{align}
        \frac{\p}{\p t}p(\bm x,t) = \mathcal{L}p(\bm x,t),
        \label{eq:FKeq}
    \end{align}
    where the Fokker-Planck operator $\mathcal{L}$ is given by
    \begin{align}
        \mathcal{L} =-\sum_{i=1}^{d}\p_{x_i}[\bmu(\bm x)]_i
        +\frac{1}{2}\sum_{i,j=1}^{d}\p_{x_i}\p_{x_j}[\bsigma(\bm x)\bsigma^T(\bm x)]_{i,j}.
    \end{align}
    Its adjoint operator, which is called the backward Kolmogorov operator, defined as
    \begin{align}
        \mathcal{L}^* = 
        \sum_{i=1}^{d}
        \bmu_i(\bm x)\p_{x_i}+
        \frac{1}{2}\sum_{i,j=1}^{d}[\bsigma(\bm x)\bsigma^T(\bm x)]_{i,j}\p_{x_i}\p_{x_j}.
        \label{eq:adjoint_op}
    \end{align}
    This operator describes the time evolution of a function $\tilde{p}(\bm x, t)$,
        \begin{align}
        \frac{\p}{\p t} \tilde{p}(\bm x,t) = \mathcal{L}^*\tilde{p}(\bm x,t),
        \label{eq:bK_eq}
    \end{align}
    which is known as the backward Kolmogorov equation.
    The function $\tilde{p}(\bm x, t)$ arises from integration by parts in the computation of a moment provided that $\bx$ takes $\bm{x}_0$ as its initial value:
    \begin{align} 
            \mathbb{E}_{\bx(t)}\left[\bx^{\bn}\right]
            & =\int_{-\infty}^{\infty} \bm{x}^{\bn} p(\boldsymbol{x}, t) d \boldsymbol{x} 
            = \int_{-\infty}^{\infty} \bm{x}^{\bn}\left(e^{\mathcal{L} t} p(\bm x,0)\right) d \boldsymbol{x} \nonumber \\ 
            & =\int_{-\infty}^{\infty} \bm{x}^{\bn}\left(e^{\mathcal{L} t} \delta(\boldsymbol{x}-\boldsymbol{x}_0)\right) d \boldsymbol{x} \nonumber \\ 
            & =\int_{-\infty}^{\infty}\left(e^{\mathcal{L}^{*} t} \bm{x}^{\bn}\right) \delta(\boldsymbol{x}-\boldsymbol{x}_0) d \boldsymbol{x}  \nonumber \\
            & =\int_{-\infty}^{\infty} \tilde{p}(\boldsymbol{x}, t) \delta(\boldsymbol{x}-\boldsymbol{x}_0) d \boldsymbol{x}  
            =\tilde{p}(\boldsymbol{x}_0, t).
        \label{eq:partial_integral}
    \end{align}
    This equation implies that once we obtain $\tilde{p}\left(\boldsymbol{x}, t\right)$ by solving Eq.~\eqref{eq:bK_eq}, the moment $\mathbb{E}_{\bx(t)}\left[\bx^{\bm{n}}\right] $ for any initial values $\bm{x}_0$ is computed by substituting $\boldsymbol{x}=\bm{x}_0$ into $\tilde{p}\left(\boldsymbol{x}, t\right)$.
        
    We may express $\tilde{p}(\boldsymbol{x}_0, t)$ as series in $\boldsymbol{x}_0$ such as
    \begin{align}
        \tilde{p}(\bm x_0, t) 
        &= \sum_{\bm{m}\in\mathbb{N}_0^d}P(\bm{m};t)\bm x_0^{\bm{m}} \nonumber \\
        &=\sum_{\bm{m}\in\mathbb{N}_0^d}P(\bm{m};t)[\bm x_0]_1^{m_1}[\bm x_0]_2^{m_2}\cdots [\bm x_0]_d^{m_d} \nonumber \\
        &=\sum_{\bm{m}\in\mathbb{N}_0^d}P(\bm{m};t)\ket{m_1,m_2,\cdots,m_{d-1},m_d}
        \label{eq:basis_expansion}
    \end{align}
    with coefficients $P(\bm{m};t)$ and the basis with braket notation \cite{ohkubo2013extended} instead of monomial of $\bm x_0$. The backward Kolmogorov equation can be reexpressed as coupled equations of the expansion coefficients as follows:
    \begin{align}
        \frac{\p}{\p t}P(\bm{n}; t) = \sum_{r=1}^{M}a_r(\bm{n}-\bm{\xi}_r)P(\bm{n}-\bm{\xi}_r;t).
        \label{eq:master_like}
    \end{align}
    Later, we will give explicit examples of this reexpression. Note that if the corresponding dual process holds the stochastic process, this equation is equivalent to the master equation associeted with $M$ events. For each event, its propensity function $a_r(\bm{n})$ and stoichiometric vector $\bm{\xi}_r\in\mathbb{Z}^d$ are determined by the backward Kolmogorov equation. 

    \subsection{Example: noisy van der Pol model $(d=2)$} \label{sec:nvdp}
    The noisy van der Pol (vdP) model is a two-dimensional SDE of the form given in Eq.~\eqref{eq:sde_highd}, with the following drift and diffusion terms:
    \begin{align}
        &\bmu(\bm x) = \begin{bmatrix}
            x_2 \\
            \epsilon x_2 (1-x_1^2) -x_1 
            \end{bmatrix} ,
        &\bm \sigma(\bm x) = \begin{bmatrix}
            \nu_{11} & 0 \\
            0 & \nu_{22}
            \end{bmatrix}.
    \end{align}
    The corresponding backward Kolmogorov operator is calculated as 
    \begin{align*}
        \mathcal{L}^* &= \begin{bmatrix}
            x_2 &
            \epsilon x_2 (1-x_1^2) -x_1 
            \end{bmatrix}
            \begin{bmatrix}
                \p_{x_1} \\
                \p_{x_2}
                \end{bmatrix} 
            +\frac{1}{2}
            \ \mathrm{tr}
                \left(
                    \begin{bmatrix}
                        \nu_{11}^2 & 0 \\
                        0 & \nu_{22}^2
                    \end{bmatrix}
                    \begin{bmatrix}
                        \p_{x_1}^2 \\
                        \p_{x_2}^2
                    \end{bmatrix}     
                \right)
        \\
        &=x_2\p_{x_1}+\epsilon x_2 (\p_{x_2}-x_1^2\p_{x_2}) -x_1\p_{x_2}+\frac{1}{2}\nu_{11}^2\p_{x_1}^2+\frac{1}{2}\nu_{22}^2\p_{x_2}^2.
    \end{align*}
    We will solve the backward Kolmogorov equation  in Sect.~\ref{sec:time_evolution}. In the bracket notation \cite{ohkubo2013extended}, applying the backward Kolmogorov operator $\mathcal{L}^*$ to the state $\ket{n_1, n_2}$, the backward Kolmogorov equation is given by
    \begin{align*}
        \p_t P(\bm{n};t) =& n_1P(n_1+1,n_2-1;t)+\epsilon n_2  P(n_1,n_2;t) \\
        &-\epsilon n_2  P(n_1-2,n_2;t) -n_2P(n_1,n_2;t)\\
        &+\frac{1}{2}\nu_{11}^2(n_1+1)(n_1+2)P(n_1+2,n_2; t)\\
        &+\frac{1}{2}\nu_{22}^2(n_2+1)(n_2+2)P(n_1,n_2+2;t ).
    \end{align*}
    The coefficients of the function $P(n_1, n_2;t)$ correspond to the propensity functions, which represent transition rates between states. The stoichiometric vectors associated with each event are denoted as follows:
    \begin{align}
            \bm{\xi}_1 = \begin{bmatrix}
                -1 \\
                1 
                \end{bmatrix},\ 
            \bm{\xi}_2 = \begin{bmatrix}
                0 \\
                0 
                \end{bmatrix},\ 
            \bm{\xi}_3 = \begin{bmatrix}
                2 \\
                0 
                \end{bmatrix},
            \nonumber \\ 
            \bm{\xi}_4 = \begin{bmatrix}
                1 \\
                -1 
                \end{bmatrix},\ 
            \bm{\xi}_5 = \begin{bmatrix}
                -2 \\
                0 
                \end{bmatrix},\ 
            \bm{\xi}_6 = \begin{bmatrix}
                0 \\
                -2 
                \end{bmatrix}.
            \label{eq:sto_vec_vanderPol}
    \end{align}
    Table~\ref{tab:noisyvanderPol} summarizes the events and their corresponding terms in the backward Kolmogorov operator.

    \begin{table}[tbp]
        \centering
        \caption{Events in the noisy vdP model.}
        \begin{tabular}{llcll}
        \hline
        Event & $a(\bm{n})$ & $\bm{\xi}$ & term/s\\
        \hline
        $X_2\to X_1$ & $n_1$ & $[-1,1]^t$ & $x_2\p_{x_1}$\\
        $X_2\to X_2 $ & $\epsilon n_2$ & $[0,0]^t$ & $\epsilon x_2\p_{x_2}$\\ 
        $X_1\to X_1 + X_2$ & $-\epsilon n_2$ & $[2,-1]^t$ & $-\epsilon x_2x_1^2\p_{x_2}$\\
        $X_1\to X_1 + X_3$ & $-n_2$ & $[1,-1]^t$ & $-x_1\p_{x_2}$\\
        $X_2\to X_2$ & $\frac{1}{2}\nu_{11}^2n_1^2$ & $[-2,0]^t$ & $\frac{1}{2}\nu_{11}^2\p_{x_1}^2$\\
        $X_2\to X_2 + X_1$ & $\frac{1}{2}\nu_{22}n_2^2$ & $[0,-2]^t$ & $\frac{1}{2}\nu_{22}^2\p_{x_2}^2$\\
        \hline
        \end{tabular}
        \label{tab:noisyvanderPol}
    \end{table}


    \subsection{Example: noisy Lotka-Volterra model $(d= 3)$}
    The Lotka-Volterra model, a famous prey-predator model, is extended to the one with diffusion as the SDE with the following drift and diffusion terms:
    \begin{align}
        &\bmu_i(\bm x) = 
            \left(\epsilon_i + \sum_{j=1}^3\mu_{ij}x_j \right)x_i
            ,
        &\bm \sigma_{ij}(\bm x) = \sigma_{i}x_i \delta_{ij},
        \label{eq:nLV_coeffs}
    \end{align}
    where $\delta_{ij}$ is Kronecker delta.
    We call the Lotka-Volterra model with the diffusion terms the noisy Lotka-Volterra (noisy LV) model. The backward Kolmogorov operator is 
    \begin{align*}
        \mathcal{L}^* 
        &= \sum_{i=1}^{3}\left(\epsilon_i + \sum_{j=1}^3\mu_{ij}x_j \right)x_i\p_{x_i}+\sum_{i,j=1}^{3}\left(\frac{1}{2}\sigma^2_{i}x^2_i \delta_{ij}\right)\p_{x_i}\p_{x_j}\\
        &= (\epsilon_1 + \mu_{11}x_1 + \mu_{12}x_2 + \mu_{13}x_3)x_1\partial_{x_1}+\frac{1}{2}\sigma_1^2x_1^2\partial^2_{x_1}\\ 
        &+ (\epsilon_2 + \mu_{21}x_2 + \mu_{22}x_2 + \mu_{23}x_3)x_2\partial_{x_2}+\frac{1}{2}\sigma_2^2x_2^2\partial^2_{x_2}\\
        &+ (\epsilon_3 + \mu_{31}x_1 + \mu_{32}x_2 + \mu_{33}x_3)x_3\partial_{x_3}+\frac{1}{2}\sigma_3^2x_3^2\partial^2_{x_3},
    \end{align*}
    and we obtain the backward Kolmogorov equation in Eq.~\eqref{eq:master_like};
    the events which happen are presented in Table~\ref{tab:noisyLVmodel_3sp}.
    
    \begin{table}[tbp]
        \centering
        \caption{Events in the noisy LV model ($d=3$).}
        \begin{tabular}{llcll}
        \hline
        Event & $a(\bm{n})$ & $\bm{\xi}$ & term/s\\
        \hline
        $X_1\to X_1$ & $\epsilon_1 n_1+\frac{1}{2}\sigma_1^2 n_1(n_1-1)$ & $[0,0,0]^t$ & $\epsilon_1x_1\p_{x_1}+\frac{1}{2}\sigma^2_{1}x^2_1\p_{x_1}^2$\\
        $X_1\to 2X_1 $ & $\mu_{11}n_1$ & $[1,0,0]^t$ & $\mu_{11}x_1x_1\p_{x_1}$\\ 
        $X_1\to X_1 + X_2$ & $\mu_{12} n_1$ & $[0,1,0]^t$ & $\mu_{12}x_2x_1\p_{x_1}$\\
        $X_1\to X_1 + X_3$ & $\mu_{13} n_1$ & $[0,0,1]^t$ & $\mu_{13}x_3x_1\p_{x_1}$\\
        \hline
        $X_2\to X_2$ & $\epsilon_2 n_2+\frac{1}{2}\sigma_2^2 n_2(n_2-1)$ & $[0,0,0]^t$ & $\epsilon_2x_2\p_{x_2}+\frac{1}{2}\sigma^2_{2}x^2_2\p_{x_2}^2$\\
        $X_2\to X_2 + X_1$ & $\mu_{21} n_2$ & $[1,0,0]^t$ & $\mu_{21}x_1x_2\p_{x_2}$\\
        $X_2\to 2X_2 $ & $\mu_{22} n_2$ & $[0,1,0]^t$ & $\mu_{22}x_2x_2\p_{x_2}$\\
        $X_2\to X_2 + X_3$ & $\mu_{23} n_2$ & $[0,0,1]^t$ & $\mu_{23}x_3x_2\p_{x_2}$\\
        \hline
        $X_3\to X_3$ & $\epsilon_3 n_3+\frac{1}{2}\sigma_3^2 n_3(n_3-1)$ & $[0,0,0]^t$ & $\epsilon_3x_3\p_{x_3}+\frac{1}{2}\sigma^2_{2}x^2_3\p_{x_3}^2$\\
        $X_3\to X_3 + X_1$ & $\mu_{31} n_3$ & $[1,0,0]^t$ & $\mu_{31}x_1x_3\p_{x_3}$\\
        $X_3\to X_3 + X_2 $ & $\mu_{32} n_3$ & $[0,1,0]^t$ & $\mu_{32}x_2x_3\p_{x_3}$\\
        $X_3\to 2X_3$ & $\mu_{33} n_3$ & $[0,0,1]^t$ & $\mu_{33}x_3x_3\p_{x_3}$\\
        \hline
        \end{tabular}
        \label{tab:noisyLVmodel_3sp}
    \end{table}

    \section{Contribution 1: Tensor-train Application to the Backward Kolmogorov Equation} \label{sec:time_evolution}
After rewriting the backward Kolmogorov equation in the TT and TTO formats, it is possible to employ numerical algorithms to solve the time-evolution equation in the TT and TTO formats. In the algorithms, the time-evolution equation is attributed to the problem of simultaneous equations; for details, see Ref.~\cite{holtz_als,dolgov2019tensor}. Here, we mainly focus on two different algorithms used in the time-evolution problem, the ALS and MALS algorithms.
\subsection{Basic formulation}
In order to solve the coupled differential equations in Eq.~\eqref{eq:master_like} in the TT format, we first denote the equations to solve in terms of tensors of order $d$ and $2d$ which represent multivariate functions, resulting in the linear system 
\begin{align}
    \frac{\p}{\p t} \mathcal{P}(t) = \mathcal{A}\ \mathcal{P}(t).
    \label{eq:CME}
\end{align}

We first introduce the TT format; when a tensor $\mathcal{X}\in\mathbb{R}^{N_1\times N_2\times\cdots\times N_d}$ of order $d$ is decomposed as the product of the third-order tensors $\mathcal{T}^{(j)}\in\mathbb{R}^{R_{j-1}\times N_j\times R_{j}} \ (j=1,\cdots,d)$:
\begin{align}
    \mathcal{X}_{i_1,i_2,\cdots,i_d} = \sum_{r_1=1}^{R_1}\cdots\sum_{r_{d-1}=1}^{R_{d-1}}\mathcal{T}^{(1)}_{1,i_1,r_1}\mathcal{T}^{(2)}_{r_1,i_2,r_2}\cdots\mathcal{T}^{(d-1)}_{r_{d-2},i_{d-1},r_{d-1}}\mathcal{T}^{(d)}_{r_{d-1},i_d,1},
    \label{eq:def_mps}
\end{align}
we say that this tensor $\mathcal{X}$ is a tensor train (\textit{TT}). The top row in Fig. \ref{fig:MPO_MPS} illustrates the TT representation. 
Here, $\mathcal{X}_{i_1,i_2,\cdots,i_d}$ in the above expression denotes the element of $\mathcal{X}$ specified with the indices $i_1,i_2,\cdots,i_d$, which are called physical indices.
In the same manner, the tensor $\mathcal{X}\in\mathbb{R}^{N_1\times M_1\times \cdots\times N_d\times M_d}$ of order $2d$ in the TT format, that is, the expression with the product of the fourth-order tensors $\mathcal{T}^{(j)}\in\mathbb{R}^{R_{j-1}\times N_j \times M_j \times R_{j}}$ $(j=1,\cdots,d)$,
\begin{align}
    \begin{split}
    &\mathcal{X}_{n_1,m_1,\cdots,n_d,m_d}\\ 
    =& \sum_{r_1=1}^{R_1}\cdots\sum_{r_{d-1}=1}^{R_{d-1}}\mathcal{T}^{(1)}_{1,n_1,m_1,r_1}\mathcal{T}^{(2)}_{r_1,n_2,m_2,r_2}\cdots\mathcal{T}^{(d-1)}_{r_{d-2},n_{d-1},m_{d-1},r_{d-1}}\mathcal{T}^{(d)}_{r_{d-1},n_d,m_d,1},
    \end{split}
    \label{eq:def_mpo}
\end{align}
is called a TT operator (\textit{TTO}). The tensors $\mathcal{T}^{(i)}\ (i=1,2,\cdots,d)$ in the products in Eqs.~\eqref{eq:def_mps} and \eqref{eq:def_mpo} are referred as \textit{TT cores}, and the indices $r_i$ in the right-hand side of Eqs.~\eqref{eq:def_mps} and \eqref{eq:def_mpo} are called \textit{TT ranks (bond dimensions)}. 

The function $P(\bn;t)$ in Eq.~\eqref{eq:master_like} is expressed as a tensor $\mathcal{P}(t)$ of order $d$; the indices of the tensor $(n_1,\cdots, n_d)$ specify the argument $\bn$ of the function, which is now truncated to $\bn\in \aleph^d\ (\aleph:=\{0,1,\cdots,N-1\})$.
    Thus, the function $P(\bn;t)\colon\aleph^d\cup[0,T]\to\mathbb{R}$ corresponds to the tensor $\mathcal{P}(t)\in\mathbb{R}^{\aleph^d}$ element-wise, as follows:
    \begin{align}
        [\mathcal{P}(t)]_{n_1,\cdots, n_d} = P(\bn;t).
        \label{eq:state_tensor}
    \end{align}

    We introduce a shift operator $ \mathcal{J}_r\in\mathbb{R}^{\aleph^d}\times\mathbb{R}^{\aleph^d}$ for the tensor $ \mathcal{P}(t)$ to represent the shift by the stoichiometric vector $ \bm{\xi}_r = (\xi_r^{(1)}, \cdots, \xi_r^{(d)})$ in the first argument of the function $ P(\bm{n} - \bm{\xi}_r; t)$ in Eq.~\eqref{eq:master_like}. This operator shifts the indices of the tensor $ \mathcal{P}$ accordingly as follows:
    \begin{align}
        \mathcal{J}_r = J(-\xi_{r}^{(1)}) \otimes \cdots \otimes J(-\xi_{r}^{(d)}),
        \label{eq:def_shift_op}
    \end{align}
    where the matrix $J(z_i) \in \mathbb{R}^{ \aleph\times \aleph}$ is a shift matrix defined as 
    \begin{align}
        [J(z_i)]_{kl}=\delta_{k+z_i, l},
        &&
        [J(-z_i)]_{kl}=\delta_{k, l+z_i},
        \end{align}
        where $\delta_{ij}$ is the Kronecker delta.

In analogy with the correspondence between the indices of a tensor and the state vector in Eq.~\eqref{eq:state_tensor}, the propensity function $a_r(\bn -\bm \xi_r)\colon\mathbb{N}_0^d\to\mathbb{R}$ is written as an element of a tensor $\mathcal{W}_r\in\mathbb{R}^{\aleph^d}$ of order $d$ specified at the index $(n_1-\xi_{r}^{(1)},\cdots ,n_d-\xi_{r}^{(d)})=:\bn -\bm \xi_r$,
    \begin{align}
        [\mathcal{W}_r]_{n_1-\xi_{r}^{(1)},\cdots ,n_d-\xi_{r}^{(d)}}= a_r(\bn -\bm \xi_r).
    \end{align}
    The tensor operator $\mathcal{A}$ in the right-hand side of Eq.~\eqref{eq:CME}
    is expressed as the \textit{product} of the stoichiometric vector $\mathcal{J}_r\in\mathbb{R}^{\aleph^d\times\aleph^d}$ and the propensity function $\mathcal{W}_r\in\mathbb{R}^{\aleph^d}$ in the tensor format
    \begin{align}
        \mathcal{A} = \sum_{r=1}^{M}\mathcal{J}_r\diag(\mathcal{W}_r),
        \label{eq:mpo_coupledODE}
    \end{align}
    where $\diag(\mathcal{W}_r)$ is the tensor of order $2d$, defined as follows:
    \begin{align}
        [\diag(\mathcal{W}_r)]_{n_1,\cdots, n_d, n_{d+1},\cdots, n_{2d} } := [\mathcal{W}_r]_{n_1,\cdots ,n_d}\delta_{n_1,n_{d+1}}\cdots\delta_{n_d,n_{2d}}.
    \end{align}

    \subsection{Example: $d\in\mathbb{N},\ M=4d-2$ case (noisy LV model)}
    The parameter $\mu_{ij}\ (1\leq i,j\leq d)$ represents the strength of interspecific competition between the $i$-th and the $j$-th species, as shown in Eq.  ~\eqref{eq:nLV_coeffs}.
    For simplicity, we assume only nearest neighbor interactions in this section;
    \begin{align}
        &\bmu_i(\bm x) = 
            \left(\epsilon_i + \sum_{j=1}^d\mu_{ij}x_j \right)x_i
            ,
        &\bm \sigma_{ij}(\bm x) = \sigma_{i}x_i \delta_{ij}
        \label{eq:nLV_coeffs_d}.
    \end{align}
    The $i$-th species has five reactions in Eq.~\eqref{eq:nLV_coeffs}: increasing with birth rate $\epsilon_i$, increasing by eating the $(i+1)$-th species as prey with death rate $\mu_{i, i-1}$,
     decreasing as prey of the $(i+1)$-th species with death rate $\mu_{i, i+1}$,
     and decreasing by intraspecific competition with death rate $\mu_{i,i}$. 
     In addition to these reactions, we consider the diffusion as well. 
    
    The above reactions reduce to four reactions in the dual process, which do not correspond with increase and decrease of particles in the SDEs.
    The propensity functions $a_{i1}(\bn), a_{i2}(\bn), a_{i3}(\bn)$ and  $a_{i4}(\bn)$ ($i=1,\cdots,d$) can be written as
    \begin{align}
    \begin{split}
        a_{i1}(\bn) &= \epsilon_i n_i+\frac{1}{2}\sigma_i^2 n_i(n_i-1),\\
        a_{i2}(\bn) &= \mu_{i,i-1}n_i,\\
        a_{i3}(\bn) &= \mu_{i,i} n_i,\\
        a_{i4}(\bn) &= \mu_{i,i+1}n_i,
    \end{split}
\end{align}
    with the corresponding stoichiometric vectors $\bm{\xi}_{ir}\in\mathbb{Z}^d\ (r=1,2,3,4)$
    \begin{align}
        \begin{split}
            \bm{\xi}_{i1}(\bn) &= \bm{o}, \\
            \bm{\xi}_{i2}(\bn) &= \bm{e}_{i-1}, \\
            \bm{\xi}_{i3}(\bn) &= \bm{e}_{i}, \\
            \bm{\xi}_{i4}(\bn) &= \bm{e}_{i+1}.
        \end{split}
    \end{align}
    Here, $\bm{e}_j$ is a unit vector which has a value of 1 at the $j$-th position, with all other elements equal to 0.
    By the definition of the shift operator in Eq.~\eqref{eq:def_shift_op}, the operator $\mathcal{A}$ in Eq.~\eqref{eq:mpo_coupledODE}
    is computed by adding the TTOs over $M$ events.
    
    \subsection{Time-evolution scheme} 
    The ALS algorithm has been applied to solve high-dimensional eigenvalue problems and linear systems with the TT format\cite{holtz_als}. In this paper, we focus on the linear system as expressed in Eq.~\eqref{eq:CME} for $\mathcal{P}(t)$, which is represented as a $d$-order tensor. Using the retraction operator, ALS reduces the original linear system to \textit{micro-equations}, which are a sequence of lower-dimensional linear equations, for optimizing each core. Then, the solution of the micro-equation is decomposed so that the core is left-orthonormal (right-orthonormal) during the `half sweep' (the `back sweep'). In Ref.~\cite{holtz_als}, QR decomposition is used in ALS, while MALS employs truncated singular value decomposition (SVD) to dynamically adjust the TT ranks, allowing for improved adaptability in the low-rank approximations.
    
    In general, MALS outperforms ALS in achieving a desired accuracy when the dimensions $N_i$, representing the maximum number of each state-space, are relatively small. This is because the computational complexity of solving micro-equations.
    However, the usage of large $N_i$ leads to high computational costs; the row and column sizes of the matrix in each of the micro-equations are multiplied by $N_{i+1}$, respectively.
    However, since we deal with relatively large $N_i$, the MALS algorithm could not be practical because of the large computational costs.
    Note that the moments are computed using the solution $\mathcal{P}(t)$ as indicated in Eq.~\eqref{eq:basis_expansion}. 
    
    Here, we propose the usage of truncated SVD in ALS to evaluate the moments in the SDEs for the following reasons: the ALS algorithm with QR decomposition would yield low-accuracy results, and the MALS algorithm needs high computational costs. In terms of accuracy and computational cost, the above proposal is optimal.

    To demonstrate this, we conduct the two numerical experiments. First, we compare truncated SVD and QR decomposition in the `sweep' procedure of the ALS algorithm. Second, we compare the MALS algorithm and the ALS algorithm with truncated SVD. 
    The ALS algorithm typically employs QR decomposition in its sweep procedure, as this method is computationally efficient but does not allow for the adaptation of TT ranks. However, we specifically chose to use truncated SVD in the ALS algorithm in the numerical experiments because it produces a more accurate solution distribution compared to QR decomposition, as demonstrated in the first experiment. On the other hand, the MALS algorithm inherently adapts the TT ranks during the sweep procedure, potentially offering further improvements in accuracy. This comparison aims to determine whether the enhanced rank adaptation in MALS compensates for its higher computational cost relative to the ALS with truncated SVD.

    \subsection{Experiment 1: Comparison of sweep procedures}
    \subsubsection{Experimental settings of experiment 1}
    To evaluate different methods within the sweep procedure of the ALS algorithm, we used the noisy vdP model described in Sect.~\ref{sec:nvdp}. The model parameters were $\epsilon=1.0$ and $\nu=0.5$.
    We compared two matrix decompositions, truncated SVD and QR decomposition, for the sweep procedures in the ALS algorithm by solving the time-evolution equation from $t=0$ to $t=0.5$ with each approach and plotting the solution distributions $P(n_1, n_2; t=0.5)$.
    The initial condition was set as $P(n_1=1, n_2=0; t=0) = 1$, with all other entries initialized to zero, which corresponds to the evaluation of $\mathbb{E}_{\bx(t)}[X_1]$.
    The maximum number of states for each variable, $N_i$, was fixed at a given value, $N$ in each experiment. In this experiment, we employ $N=20$ and 50. Additionally, the TT ranks, denoted as $R$, were kept identical for all TT cores in both decomposition methods to ensure a fair comparison.

    As a reference, we used the solution obtained by the conventional CN method. Unlike the ALS algorithm, which approximates the solution using the TT format to achieve significant computational savings, the CN method computes the solution directly without employing any low-rank approximation. Although the CN method is computationally more expensive, it provides a reference solution that is free from the errors introduced by the TT approximation. Note that we tried several other numerical experimental settings and obtained the same consequences. Therefore, the results of the numerical experiments with the above settings are shown below.

    \subsubsection{Numerical results of experiment 1}
\begin{figure*}[bpt]
        \centering
        \includegraphics[width=0.7\linewidth]{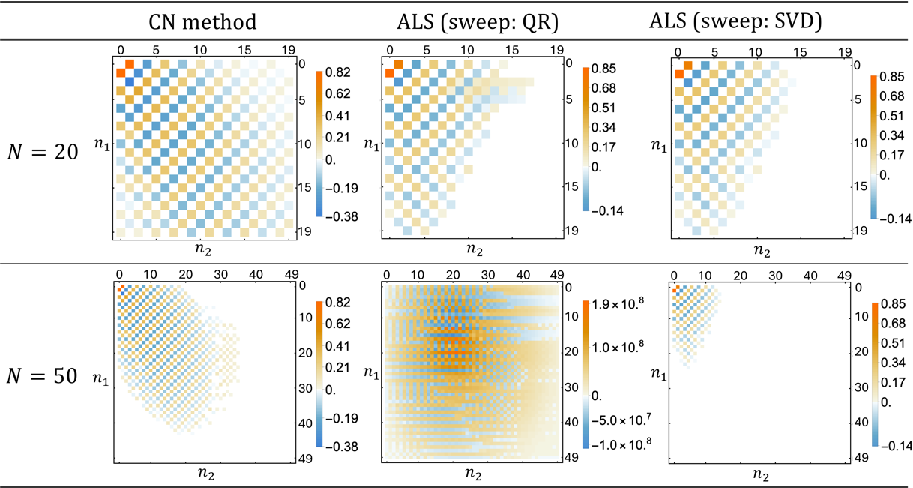}
        \caption{(Color online) Comparison of the solution distributions $P(n_1, n_2;t=0.5)$ obtained using the CN method, ALS with QR decomposition, and ALS with truncated SVD for 
        $N=20$ (upper panels) and $N=50$ (lower panels). The vertical axis represents the first index $n_1$, and the horizontal axis represents the second index $n_2$ of the solution $P(n_1, n_2;t=0.5)$. The color of each cell indicates the value of the solution, as shown in the legends. 
        Both decompositions (QR and truncated SVD) preserve the pre-set TT rank ($R=4$ for $N=20$ and $R=5$ for $N=50$).}
        \label{fig:sol_distributions}
    \end{figure*}

    Figure~\ref{fig:sol_distributions} illustrates the solution distributions $P(n_1, n_2;t=0.5)$ obtained using each method for $N=20$ and $50$. The vertical and horizontal axes represent the indices $n_1$ and $n_2$, respectively, while the color of each cell indicates the value of the solution.

    The leftmost panels in Fig.~\ref{fig:sol_distributions} show the reference solutions obtained by the CN method. The upper-left and lower-left panel correspond to $N=20$ and $N=50$, respectively. As expected, the CN method yielded the most accurate solution due to its avoidance of truncation errors, but it was computationally expensive. 
    The reference solutions demonstrate that the noisy vdP model produces a distinctive checkered pattern as it evolves over time. 

    QR decomposition, commonly used for the sweep procedure in the ALS algorithm, showed varying performance depending on $N$. 
    QR decomposition (upper-center panel in Fig.~\ref{fig:sol_distributions}) was able to reproduce the checkered pattern almost as effectively as SVD (upper-right panel). However, for $N=50$, the performance of QR decomposition shows a notable decline in performance. As shown in the lower-center panel of Fig.~\ref{fig:sol_distributions}, QR decomposition failed to reproduce the checkered pattern, resulting in divergence, which is evident from the scale of its color bar, and caused the peak of the solution to shift toward the central region. 

    In contrast, truncated SVD preserved the checkered pattern for both $N=20$ and $50$, as seen in the upper-right and lower-right panels of Fig.~\ref{fig:sol_distributions}, respectively. Note that the range of the area where the checkered pattern was reproduced pattern remained unchanged regardless of $N$. For $N=50$, when compared to the solution obtained by the CN method, the checkered pattern produced by SVD covered a smaller region, indicating limitations in fully capturing the distribution accurately as the CN method due to the TT approximation.
    
    The better performance of truncated SVD over QR decomposition in preserving the checkered pattern may be attributed to differences in their matrix decomposition properties. In the ALS algorithm, each TT core is updated using an orthogonal matrix, and subsequent cores are not influenced by the remaining decomposed matrix. However, an orthogonal matrix produced by truncated SVD provides an enhanced ability to express the solution than that generated by QR decomposition. The truncated SVD decomposes a matrix into $U\Sigma V= U \widetilde{V}$, where $U$ and $V$ are orthogonal matrices, and $\Sigma$ is a diagonal matrix. In contrast, QR decomposition produces an orthogonal matrix and an upper triangular matrix. The additional `orthogonality' in the matrix decomposed by truncated SVD likely contributes to its improved ability to represent the solution accurately.



    \subsection{Experiment 2: comparison of algorithms}
    \subsubsection{Experimental settings of experiment 2}
    To evaluate the performance of the ALS algorithm with truncated SVD (hereafter referred to as simply ALS) and the MALS algorithms, we conducted experiments using the noisy LV model for $d=4$. Two sets of parameter configurations were considered. For the first case, the strengths of the interactions in Eq.~\eqref{eq:nLV_coeffs_d} were randomly chosen as follows:
    \begin{align*}
        &[\mu_{ij}]=\begin{bmatrix}
            -0.79 & 0.07 & 0 & 0 \\
            -0.43 & 0.09 & 0.18 & 0 \\
            0 & -0.6 & 0.93 & -0.19 \\
            0 & 0 & 0.98 & -0.03 \\
            \end{bmatrix}, \\
        &\bm\epsilon=\begin{bmatrix}
                0.37 & 0.99 & 0.52 & -0.93
            \end{bmatrix}, 
    \end{align*}
    with no diffusion, that is, $\bm \sigma = \bm 0$. For the second case, the parameters were set as:
    \begin{align}
        &[\mu_{ij}]=\begin{bmatrix}
            0 & -0.12 & 0 & 0 \\
            0.12 & 0 & -0.14 & 0 \\
            0 & 0.08 & 0 & 0.14 \\
            0 & 0 & 0.06 & 0 \\
            \end{bmatrix}, \\
        &\bm\epsilon=\begin{bmatrix}
                0.6 & 0.4 & 0.2 & -0.42
            \end{bmatrix}, 
        \label{eq:params_oscillation}
    \end{align}
    and $\bm \sigma = \bm 0$. This parameter setting led to the oscillation in the state variables, $\bx$.
    The initial values of the variables were set as $\bm{x}_0 = \bm 1 = [1.0, 1.0, 1.0, 1.0]$ for the first case to compute the moments. For the second case, we computed the moments with the initial value $\bm{x}_0 = \bm 1 = [1.0, 1.0, 1.0, 1.0]$ and $\bm{x}_0 =  5\cdot \bm 1= [5.0, 5.0, 5.0, 5.0]$. Both the ALS and MALS algorithms employed $N=10$ and $R=10$. For each algorithm, we computed the prediction error of the moment $\mathbb{E}_{\bx(t)}[\mathrm{X}_3]$, obtained through the dual process method in Sect.~\ref{sec:method}, and assessed the results based on prediction errors. 
    The reference value was computed by solving the coupled differential equations 
    \begin{align}
        \frac{d \bm X}{dt} = \bmu(\bm X),
        \label{eq:alsmals_diff_eq}
    \end{align}
        which are deterministic since there is no diffusion, using the \texttt{scipy.integrate.odeint} function in Python. 



    \subsubsection{Numerical results of experiment 2}
\begin{figure}[bt]
        \centering
        \includegraphics[width=0.85\linewidth]{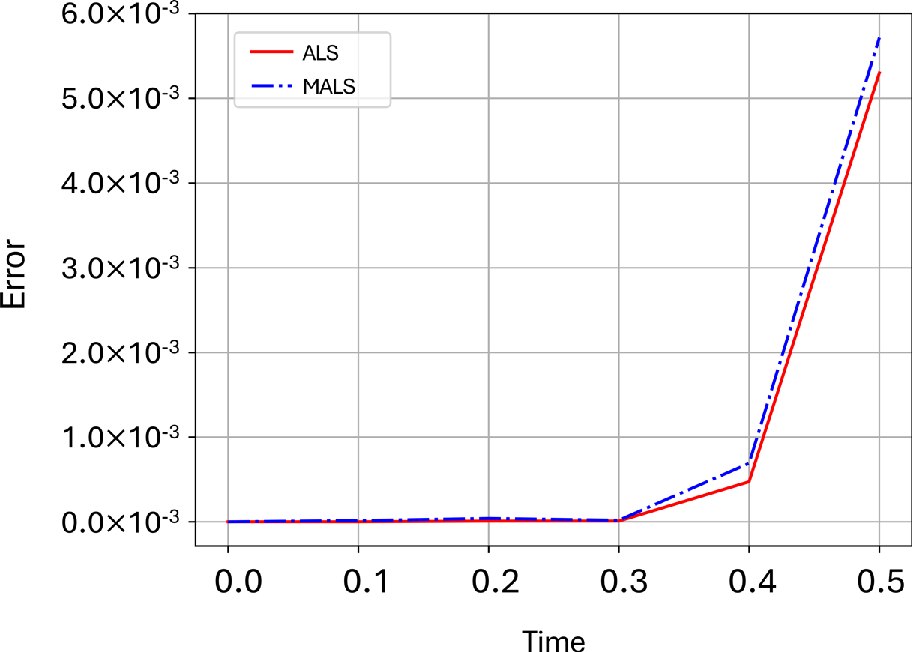}
        \caption{(Color online) Prediction error of $\mathbb{E}_{\bx(t)}[\mathrm{X}_3]$ with $\bm x_0 = \bm 1$ over time for the ALS algorithm with truncated SVD and the MALS algorithm. The error is computed as the difference between the moments obtained by each method and the reference solution provided by a scipy library function on Python. ALS achieves smaller errors than MALS, especially at later time points.}
        \label{fig:ALSMALSerror}
    \end{figure}

    To compare the performance of the ALS algorithm with SVD and the MALS algorithm, we first evaluated their prediction errors. 
    Figure~\ref{fig:ALSMALSerror} shows the prediction errors over time for both algorithms. The errors were computed as the difference between the moments obtained by each algorithm and the reference. In this time range, while ALS consistently achieved slightly smaller errors than MALS, the overall differences between the two algorithms were relatively small. This suggests that, under the current experimental settings, both algorithms perform similarly in terms of accuracy.

    In terms of computational time, the ALS algorithm was the fastest method to solve the coupled differential equation in Eq.~\eqref{eq:alsmals_diff_eq}; for $N=5$, the CN method, ALS, and MALS took 30 seconds, 0.4 seconds, and  0.8 seconds, respectively, under the same computational resources. This highlights the computational efficiency of the TT format. For the case when $N=10$, ALS took only 3 seconds while MALS took 90 seconds. The increased time for MALS arises from solving each `micro-equation', which involves a $(RN)^2$-dimensional linear system, compared to the $R^2N$-dimensional linear system solved by ALS. As $N$ increases, the computational advantage of ALS becomes more pronounced.

    Given the limited differences observed in prediction errors, we examine the TT cores constructing the solution to identify any distinctions resulting from the differences of the two algorithms. The primary difference between the two algorithms in this experiment is that MALS updates two adjacent TT cores simultaneously during each sweep, while ALS updates one core during each sweep. Notably, MALS does not adapt the TT rank when contracting two adjacent cores; it retains only the first ten singular values, as does ALS, since we set $R=10$. To investigate whether these algorithmic differences are reflected in the solution, we compared the distributions of each TT core constructing the solution at $t=0.5$.

\begin{figure}[tbp]
        \centering
        \includegraphics[width=0.9\linewidth]{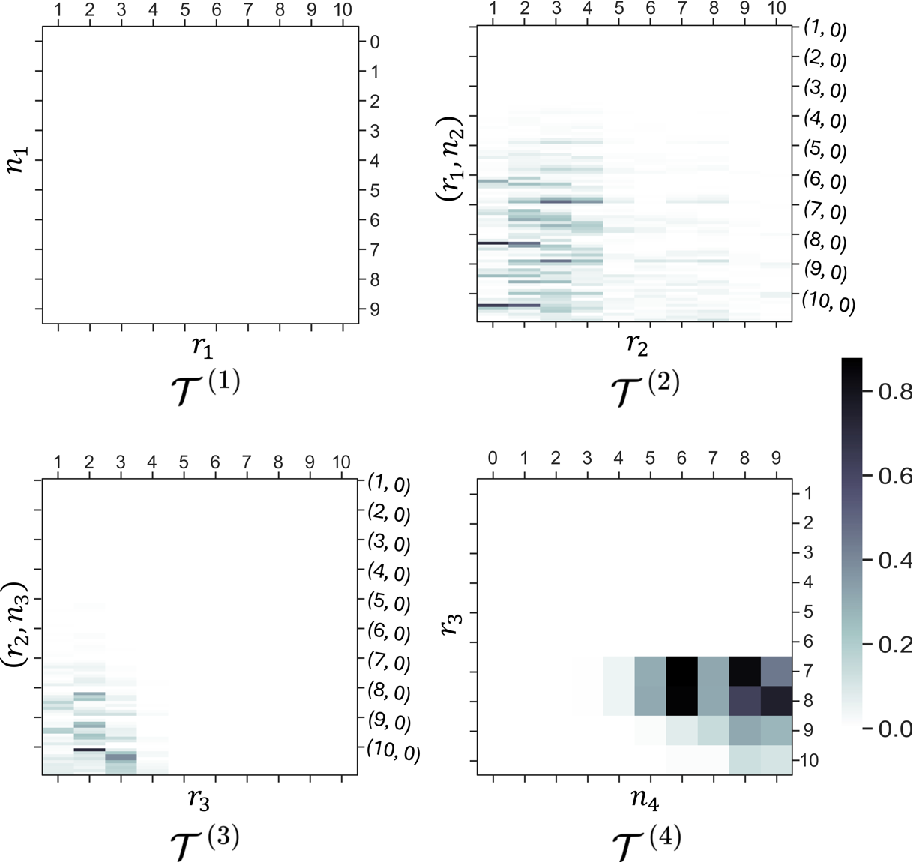}
        \caption{(Color online) Absolute differences in the solution distributions at $t=0.5$ of the TT cores produced by each algorithm, shown as matricized representations with dimensions labeled on the vertical and horizontal axes of each panel. While the first TT core $\mathcal{T}^{(1)}$ is nearly identical for both methods, significant differences emerge in the second TT core $\mathcal{T}^{(2)}$, particularly in regions where the TT rank $r_1$ exceeds five, and propagate to the TT cores $\mathcal{T}^{(3)}$ and $\mathcal{T}^{(4)}$ .}
        \label{fig:ALSMALSdiff}
    \end{figure}

    Figure~\ref{fig:ALSMALSdiff} shows the absolute differences in the solution distributions of each TT core produced by each algorithm, visualized as matricized representations.  The dimensions of the matrices are specified on the vertical and horizontal axes of each panel. The first TT core $\mathcal{T}^{(1)}$ showed minimal differences between the two algorithms, indicating that the first TT core is not significantly influenced by the differences in the sweep procedure. However, differences were observed in the second core $\mathcal{T}^{(2)}$, particularly in regions where the TT rank $r_1$ (linking the two TT cores $\mathcal{T}^{(1)}$ and $\mathcal{T}^{(2)}$) exceeded five. Darker regions in Fig.~\ref{fig:ALSMALSdiff} indicate greater absolute differences between the TT core values produced by the two algorithms. 
    These differences propagated to the TT cores $\mathcal{T}^{(3)}$ and $\mathcal{T}^{(4)}$.

\begin{figure}[t]
            \centering
            \includegraphics[width=0.85\linewidth]{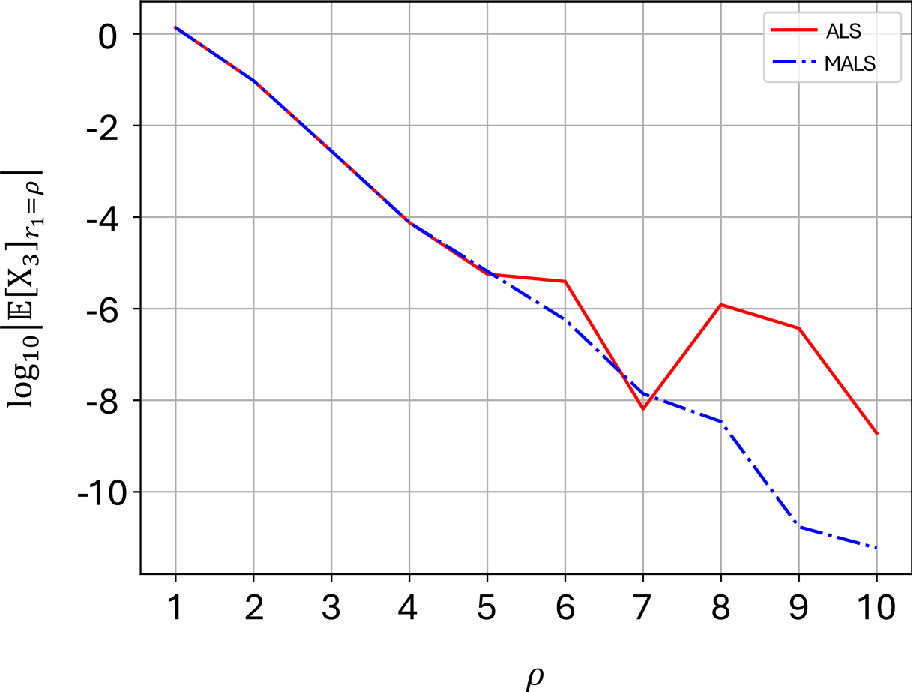}
            \caption{(Color online) Logarithm of the absolute values of the computed moment at $t=0.5$ plotted for each integer $\rho\ (1\leq \rho \leq R_1=10)$, with points connected by lines for visualization, for the ALS (solid line) and MALS (dot-dashed line) algorithms.}
            \label{fig:ALSMALSranks}
        \end{figure}

    Based on the results shown in Fig.~\ref{fig:ALSMALSdiff}, we further examined the contribution of $r_1$ to the accuracy of computed moments. Here, we define $\rho$ as a fixed value of the first TT rank $r_1$, allowing us to analyze its impact on the solution $P(\bm{m};t)$. We computed the moment $\mathbb{E}_{\bx(t)}[\mathrm{X}_3]$ using different values of $\rho$, where $1 \leq \rho \leq R$ ($R = 10$), as shown in Fig.~\ref{fig:ALSMALSranks}. Instead of summing over all possible $r_1$ values in the TT representation, we restricted the expansion to $\rho$ while summing over the remaining TT ranks $r_2, r_3, \dots, r_{d-1}$. This approach isolates the effect of $r_1$ on the computed moments and provides insight into its role in determining the final accuracy.
    The horizontal axis in Fig.~\ref{fig:ALSMALSranks} represents $\rho$. Up to $\rho = 5$, the results from both ALS and MALS were identical, as indicated by the overlapping lines. This behavior is consistent with the results in Fig.~\ref{fig:ALSMALSdiff}, where differences in the second TT cores $\mathcal{T}^{(2)}$ emerged only for $r_1 > 5$. 
    In the range $5 < \rho \leq 10$, while the values in MALS decrease exponentially, ALS does not exhibit the same behavior. However, this difference becomes negligible when the moments were cumulatively summed over $r_1$ up to $R=10$, as shown in Fig.~\ref{fig:ALSMALSerror}.


\begin{figure}[tbp]
        \centering
        \includegraphics[width=0.75\linewidth]{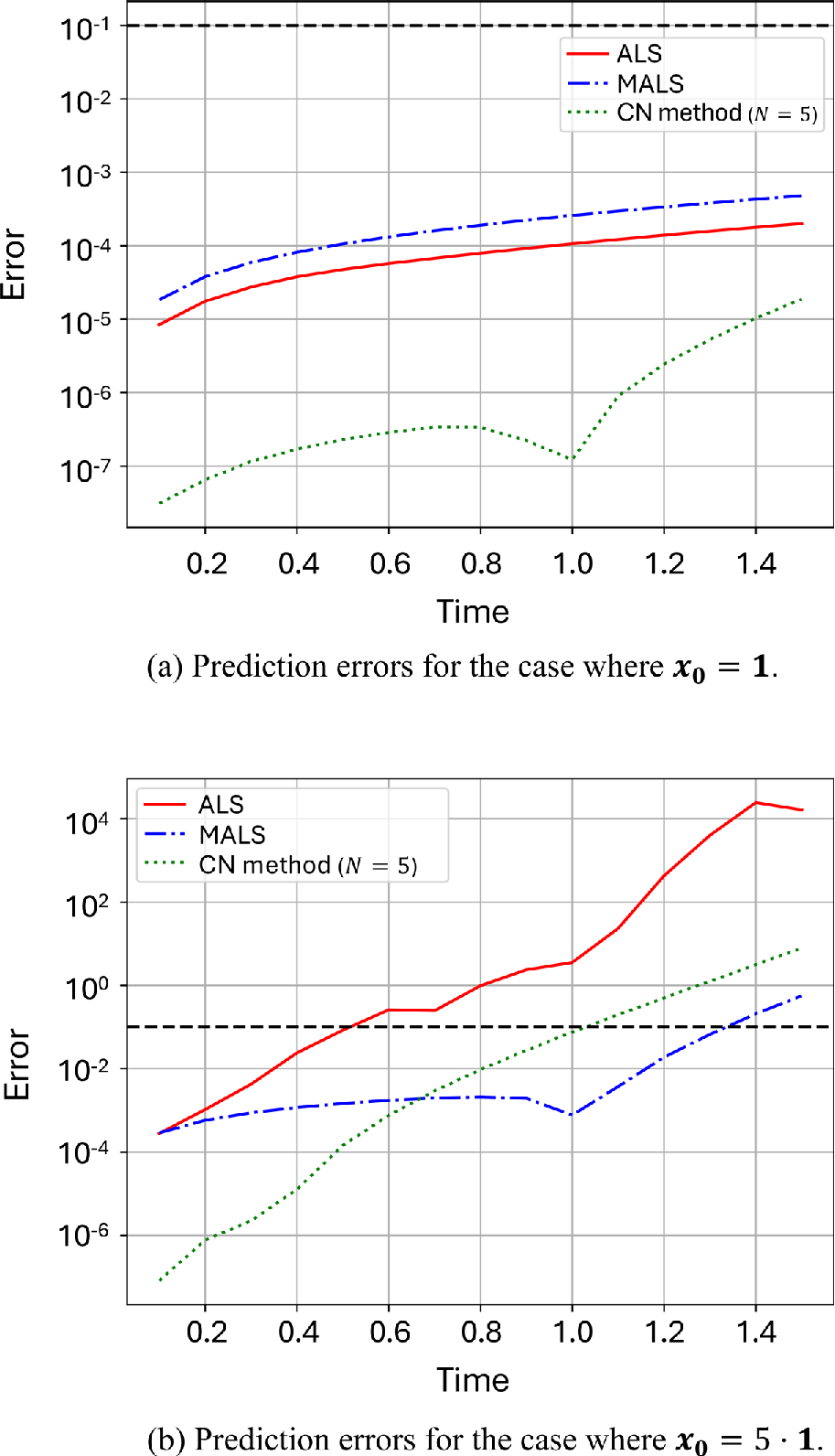}
        \caption{(Color online) Prediction errors between the reference and the moments over time for the ALS (solid line) and the MALS (dot-dashed line) algorithms with $N=10$. The initial values were set as $\bm x_0=\bm 1$ and $\bm x_0=5\cdot\bm 1$ for panels (a) and (b), respectively. As a comparison, the errors are plotted with the method solving the coupled ODE via the dual process with the conventional CN method with $N=5$ (dotted line). }
        \label{fig:ALSMALSwave_error_all}
    \end{figure}

    To further investigate the algorithmic differences, we conducted simulations using the parameter set in Eq.~\eqref{eq:params_oscillation} and the two types of initial values  $\bm x_0 = \bm 1$ and $5\cdot\bm 1$, where $\bm 1$ denotes the all entries of the $d$-dimensional vector are one. The prediction errors obtained from each algorithm with $N=10$ and $R=10$ are shown in Figs.~\ref{fig:ALSMALSwave_error_all}(a) and \ref{fig:ALSMALSwave_error_all}(b), corresponding to the initial values $\bm x_0 = \bm 1$ and $5\cdot\bm 1$, respectively. For comparison, the results of the conventional CN method with $N = 5$ are also shown as dotted lines. In both panels, the CN method achieves the smallest error, as it does not require rank approximation.
    Similar to Fig.~\ref{fig:ALSMALSerror}, when the initial values are not large enough, the two algorithms employing the TT format exhibit the similar performance, and ALS achieves slightly better accuracy than MALS. However, for large initial values such as $\bm x_0 = 5\cdot \bm 1$,   
    the two algorithms exhibit different performance behaviors. 
    The prediction errors of ALS, shown in the solid line, constantly increasing from $t \sim 0$, whereas the errors in MALS remain small until approximately $t \sim 1.1$.

    We hypothesized that this behavior results from the initial values of the variables, $\bm{x}_0$, being greater than 1.0, in combination with the parameter settings. As indicated in Eq.~\eqref{eq:basis_expansion}, when $\bm{x}_0 > 1$, larger values of $m_i$ $(i=1,2,\dots,d)$ contribute more significantly to the computed moments due to their stronger influence on the solution $P(\bm{m};t)$. To test this hypothesis, we computed the moments using a truncated solution, where only values in the range $1 \leq m_i \leq 5$ were considered, and compared the resulting errors with the reference solution.
    
\begin{figure}[ht]
        \centering
        \includegraphics[width=0.85\linewidth]{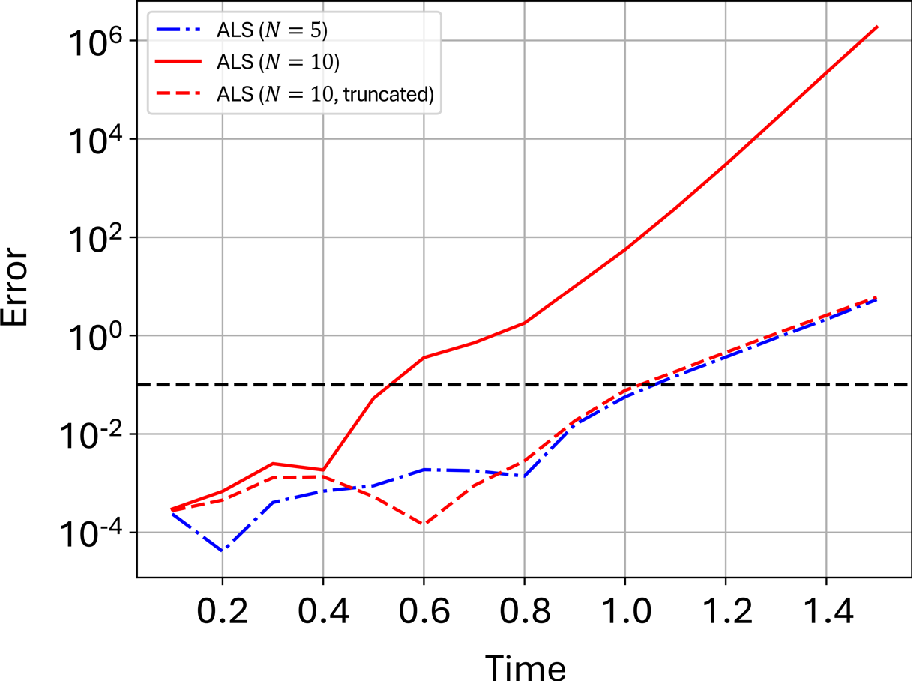}
        \caption{(Color online) Prediction errors between the reference and the computed moments over time, plotted with points connected by lines for visualization, for the ALS algorithm. Solid and dot-dashed lines indicate the cases for $ N = 10$ and $ N = 5 $, respectively. The dashed line represents the ALS algorithm with 
        $ N = 10 $, but the moment was computed using only up to $ N = 5 $, and this truncated moment was used to compute the error.
        }
        \label{fig:ALSMALSpicked}
    \end{figure}

    Figure~\ref{fig:ALSMALSpicked} presents the results. The dashed line represents the error between the reference solution and the moments computed using the truncated solution $P(\bm{m}; t)$ with $1 \leq m_i \leq 5$. This error closely aligns with the dot-dashed line for $N = 5$, supporting our hypothesis that larger values of $m_i$ are responsible for the rapid increase in the prediction error by the ALS algorithm.

    \section{Contribution 2: Ordering of TT Cores} \label{sec:order_results}
    In the previous section, we solved the coupled differential equations derived from the corresponding SDE using the ALS and MALS algorithms and compared the results in simple systems. For simple systems, such as those with nearest-neighbor interactions, the $i$-th TT cores of both the TT and the TTO, which represent the solution and the operator of the coupled equations, respectively, directly correspond to the $i$-th variable in the SDE.
    However, the interactions are often more intricate for the coupled differential equations derived from the SDE, so that the optimal ordering of the TT cores becomes non-trivial.

    We first examine how the permuted orderings of the TT cores affect the moments. 
    Here, the results we present remain consistent even when noise $\bsigma$ in Eq.~\eqref{eq:sde_highd} is introduced into the model. Hence, for clarity, we present the results without noise, as the observed tendency is more clearly distinguishable in this context.
    Then, we evaluate the relative errors of prediction results of the proposed method against those of direct numerical integration using a time step of $\Delta t = 1.0 \times 10^{-3}$.

    \subsection{Experimental settings}
    We employed the $d$-dimensional noisy LV model for this experiment.
    To systematically explore different orderings of the TT cores, we introduce a permutation function $\sigma\in {S\!}_d$, where $S_{\!d}$ is symmetric group of degree $d$. The permuted TT cores are then expressed as: 
    \begin{align*}
    \left\{\widetilde{\mathcal{T}}^{(i)}\right\}_{i=1}^d
    =\left\{ \mathcal{T}^{(\sigma(i))}\right\}_{i=1}^d.
    \end{align*}
    For example, in the case of $d=4$, the solution of Eq.~\eqref{eq:CME}, $\mathcal{P}(t)\in\mathbb{R}^{\aleph^4}$, is represented in the TT format as:
    \begin{align*}
        \sum_{r_1,r_2,r_3}\mathcal{T}^{(1)}_{1,n_1,r_1}\mathcal{T}^{(2)}_{r_1,n_2,r_2}\mathcal{T}^{(3)}_{r_2,n_3,r_3}\mathcal{T}^{(4)}_{r_3,n_4,1}.
    \end{align*}
    In the standard ordering, the $i$-th TT core corresponds to the $i$-th species. However, in this study, we explore permuted TT cores, meaning that the physical indices are reordered in the TT decomposition. We next explain the details of systematically generating and implementing different orderings of the TT cores in our experiments.
    
    The parameters in the SDE of the permuted system are derived using the orthogonality of the permutation matrix, which represents the permutation $\sigma$. 
    The SDE of the noisy LV model can be rewritten as 
    \begin{align}
    \begin{split}
        &\left((\bm\epsilon I)^T \bm{x} + (\bm{x} I)^T M\bm{x} \right)\diff t 
        +\left(( \bsigma I)^T\bm{x}\right) \diff \bW(t)\\
        =&\left((\bm\epsilon I)^TP^TP \bm{x} + (\bm{x} I)^T P^TPMP^TP\bm{x} \right)\diff t 
        +\left(( \bsigma I)^TP^TP\bm{x}\right) \diff \bW(t)\\
        =&\left((\bm{\tilde\epsilon} I)^T \bm{\tilde{x}} + (\bm{\tilde{x}} I)^T \tilde{M}\bm{\tilde{x}} \right)\diff t 
        +\left(( \tilde\bsigma I)^T\bm{\tilde{x}}\right) \diff \bW(t),
    \end{split}
    \end{align}
    with $\widetilde{\bm{\epsilon}}=P\bm\epsilon, \widetilde\bsigma = P \bsigma, 
    \widetilde{M}=PMP^T$ and $\bm{\tilde{x}} = P \bm x$.
    To consider the equations with the permuted TT cores, we solve the equation with these new parameters, 
    $\widetilde{\bm{\epsilon}},\ \widetilde\bsigma,\ \widetilde{M}=[\tilde\mu_{ij}]$.

    We conducted experiments starting with the cascade process for $d=4$ and $d=5$ with the following model parameters,
    \begin{align}
        \mu_{ij}=\begin{cases}
            \ 1.3 &(i-1 \leq j\leq i+1)\\
            \ 0 & \mathrm{otherwise}
            \end{cases}, &
        &[\bm\epsilon]_i= 0.5,
        \label{eq:params_dim4_cascade}
    \end{align}
    with no diffusion, that is, $\bm \sigma = \bm 0$.
    Additionally, we also examined cases where the strengths of the two-body interactions are randomly chosen among $\{0.0, 0.3, 0.6,0.9,1.2\}$ for $d=5$. Here, we use the following parameter set:
    \begin{align}
        [\mu_{ij}]= \begin{bmatrix}0.9 & 1.2 & 1.2 & 0.3 & 0.9\\
            0.3 & 0.3 & 0.6 & 0.9 & 0.3\\
            0.6 & 0.9 & 0.6 & 0 & 1.2\\
            0 & 0 & 0.6 & 0.3 & 1.2\\
            0.6 & 0.6 & 0.6 & 0.9 & 0.6
        \end{bmatrix}, &
        &[\bm\epsilon]_i= 0.5,
        \label{eq:params_dim5_rand}
    \end{align}
    with no diffusion, that is, $\bm \sigma = \bm 0$.
    For both models, we computed all the first and second moments, that is, the 14 and 20 different moments for $d=4$ and 5, respectively. We computed the moments up to $t=0.2$. Initial values of $\bm x$ were set $[1.1,\ 1.1,\ 1.1,\ 1.1]$ for $d=4$ and $[1.1,\ 1.1,\ 1.1,\ 1.1,\ 1.1]$ for $d=5$. 

    Each moment was computed for $d!$ different orderings of TT cores. The equations were solved using the ALS algorithm with SVD as the sweep procedure, as discussed in Sect.~\ref{sec:time_evolution}. The TT rank was set to $R=5$ among all the TT cores, and the maximum number of state space $N$ was 20. This setup leads to middle TT cores with dimensions $5\times 20\times 5$ and TT cores at both ends with dimensions $1\times 20\times 5$.
    \subsection{Numerical results}
\begin{figure}[tbp]
        \centering
        \includegraphics[width=0.85\linewidth]{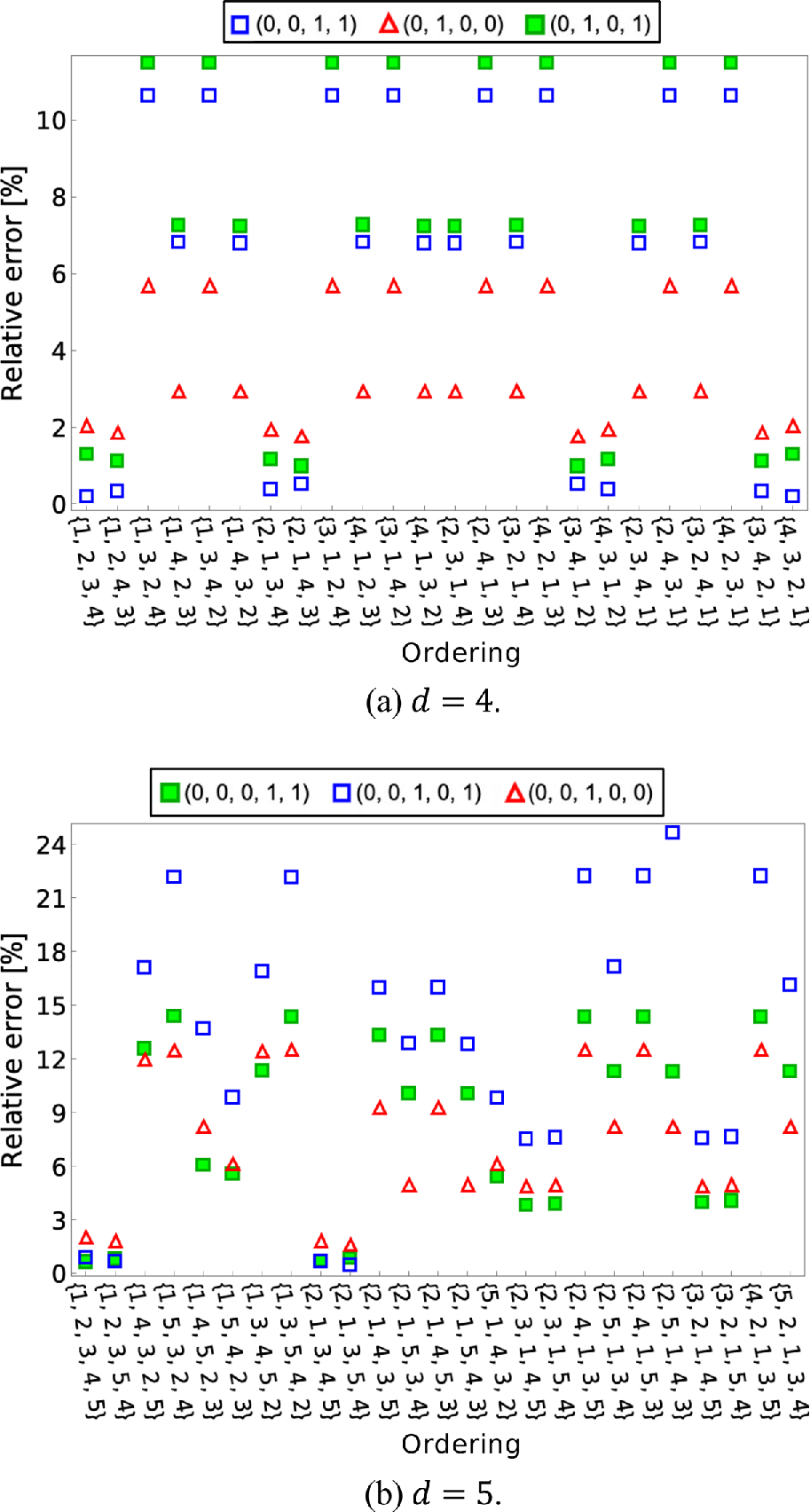}
        \caption{(Color online) Relative errors of the computed moments for different orderings of the TT cores in the noisy LV model with the identical strengths of interaction $\mu_{ij}$, shown for $d=4$ and $5$ in panels (a) and (b), respectively. The horizontal axis represents the orderings of TT cores, and the vertical axis shows the relative error in percentage. (a) Empty squares, empty triangles, and filled squares correspond to different moments $\mathbb{E}_{\bx(t)}[\bx^{\bm n}]$: $\bm n=(0,0,1,1)$, $\bm n=(0,1,0,0)$, and $\bm n=(0,1,0,1)$, respectively. (b) Empty squares, empty triangles, and filled squares correspond to different moments $\mathbb{E}_{\bx(t)}[\bx^{\bm n}]$: $\bm n=(0,0,1,0,1)$, $\bm n=(0,0,1,0,0)$, and $\bm n=(0,0,0,1,1)$, respectively.}
        \label{fig:relerror_order_dim45}
    \end{figure}

    Figure~\ref{fig:relerror_order_dim45}(a) shows some relative errors $\mathbb{E}_{\bx(t)}[\bx^{\bm n}]$ for various orderings of the TT cores, with moments $\mathbb{E}_{\bx(t)}[\rm{X}_3\rm{X}_4]$ represented by empty squares, $\mathbb{E}_{\bx(t)}[\rm{X}_2]$ represented by empty triangles, and $\mathbb{E}_{\bx(t)}[\rm{X}_2\rm{X}_4]$ represented by filled squares. For all the moments, the relative errors depend on the ordering of the TT cores. To compute the moments, we solve the couple differential equations in Eq.~\eqref{eq:CME}, starting from an initial state $\bm n$ that corresponds to the exponents of the moments $\mathbb{E}_{\bx(t)}[\bx^{\bm n}]$. 
    
    The nearest-neighbor interactions with identical strengths generally lead to consistent error patterns across different orderings. As shown in Fig.~\ref{fig:relerror_order_dim45}(a), the relative errors show the reflection symmetry in the orderings of the TT cores: the ordering $\{\sigma(1), \sigma(2),\sigma(3),\sigma(4)\}$ produces the same relative error as its reverse ordering, $\{\sigma(4), \sigma(3),\sigma(2),\sigma(1)\}$. However, certain orderings consistently yield smaller relative errors than others. 
    Some first moments exhibit less variance in relative errors compared to the second moments (not shown in figures). For the second moments (empty squares and filled squares in Fig.~\ref{fig:relerror_order_dim45}(a)), the error variation is more complex, likely due to a greater variety of interactions between species involved during the solution process. 

    The relative errors for the moments in Fig.~\ref{fig:relerror_order_dim45}(a) are grouped into three distinct layers. The gaps between these layers differ for each moment, as indicated by the legends. For example, moments such as $\mathbb{E}_{\bx(t)}[\rm X_3]$ (not shown in Fig.~\ref{fig:relerror_order_dim45}(a)) and $\mathbb{E}_{\bx(t)}[\rm X_3\rm X_4]$ shown in Fig.~\ref{fig:relerror_order_dim45}(a) exhibit similar layer structures, differing only in the relative magnitudes of the gaps between layers.

    For most cases, except for $\bm n = (1,0,0,0)$, $\bm n = (2,0,0,0)$ and $\bm n = (0,0,0,1)$, the orderings of the TT cores with adjacent numbers in the first two or last two pairs (e.g., $\{1,2,3,4\}$, $\{1,2,4,3\}$, $\{2,1,3,4\}$, and $\{2,1,4,3\}$) produce the smallest relative errors, as illustrated in Fig.~\ref{fig:relerror_order_dim45}(a). The orderings containing only one adjacent pair (e.g., $\{2,3\}$) result in the second-smallest relative errors. The remaining orderings yield the largeest relative errors.

    We also conducted experiments for $d = 5$ with the nearest-neighbor interactions of identical strengths, and the results are shown in Fig.~\ref{fig:relerror_order_dim45}(b). 

    For $d = 5$, there are $5! = 120$ possible orderings of the TT cores. For simplicity, Fig.~\ref{fig:relerror_order_dim45}(b) displays only a subset of these orderings. 
    The moments shown in Fig.~\ref{fig:relerror_order_dim45}(b) are $\mathbb{E}_{\bx(t)}[\mathrm{X}_4\mathrm{X}_5]$ (filled squares), $\mathbb{E}_{\bx(t)}[\mathrm{X}_3]$ (empty squares), and $\mathbb{E}_{\bx(t)}[\mathrm{X}_3\mathrm{X}_5]$ (empty triangles).
    The reflection symmetry for the orderings of the TT cores ensures that the relative error for a particular ordering is the same as that for its reverse ordering as well in this case. 

    The patterns observed for $d = 4$ also hold for $d = 5$. Specifically, the orderings of the TT cores containing three adjacent pairs (e.g., $\{1,2\}$, $\{2,3\}$, and $\{5,4\}$ in $\{1,2,3,5,4\}$) produce the smallest relative errors. Similarly, the orderings containing two adjacent pair (e.g., $\{2,3\}$ and $\{4,5\}$ in $\{2,3,1,4,5\}$) result in the second-smallest relative errors, while all other orderings yield the largeest relative errors.
    These results further confirm that orderings of the TT cores maintaining adjacency in nearest-neighbor interactions tend to yield smaller relative errors, emphasizing the importance of considering interaction structures when selecting orderings of the TT cores.

\begin{figure}[tbp]
        \centering
        \includegraphics[width=0.8\linewidth]{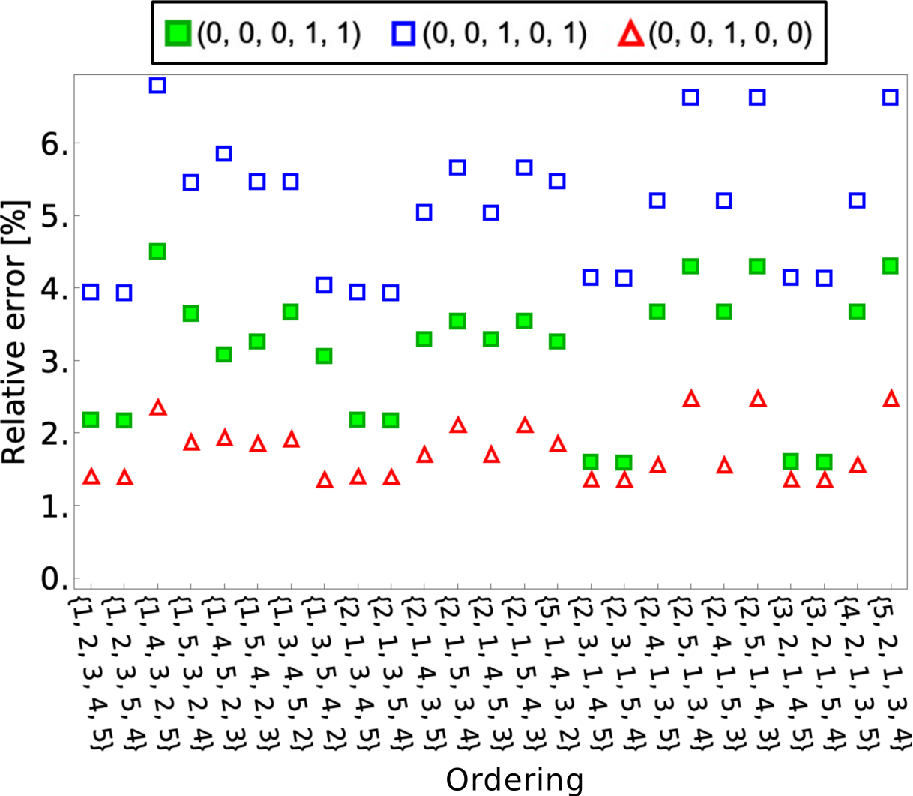}
        \caption{(Color online) Relative errors of the computed moments for different orderings of the TT cores in the noisy LV model with the randomly chosen interactions in Eq.~\eqref{eq:params_dim5_rand} with $d=5$. Empty squares (blue), empty triangles (red), and filled squares (green) correspond to different moments $\mathbb{E}_{\bx(t)}[\bx^{\bm n}]$: $\bm n=(0,0,1,0,1)$, $\bm n=(0,0,1,0,0)$, and $\bm n=(0,0,0,1,1)$, respectively. The horizontal axis represents the orderings of TT cores, and the vertical axis shows the relative error in percentage.}
        \label{fig:relerror_order_dim5_rand}
    \end{figure}

    We also conducted simulations using the randomly chosen two-body interaction strengths as specified in Eq.~\eqref{eq:params_dim5_rand}. Figure~\ref{fig:relerror_order_dim5_rand} presents the results for a subset of 25 orderings of the TT cores, selected from the $5!$ total possible orderings. Compared to Fig.~\ref{fig:relerror_order_dim45}(b), the ranges of relative errors for all moments are smaller with this parameter set. Notably, the first moments show less sensitivity to orderings. This suggests that the effect of orderings diminish when the interaction strengths are randomized, especially for the first moments. 
    Despite the random interaction strengths, the general patterns observed for the nearest-neighbor interactions remain broadly consistent. For instance, orderings of the TT cores with more adjacent pairs still tend to produce smaller relative errors, while those with fewer or no adjacent pairs result in largeer errors. As shown in Fig.~\ref{fig:relerror_order_dim5_rand}, when interaction strengths are randomized, the gap between the orderings with the smallest relative errors and those with the second-smallest relative errors is less pronounced compared to Fig.~\ref{fig:relerror_order_dim45}(b).

    To fully capture the overall patterns that determine which orderings of the TT cores lead to smaller or largeer relative errors, it is essential to move beyond a purely local perspective that considers individual TT core pairs or triplets of interactions. A more comprehensive framework is needed to systematically evaluate the relationship between orderings of the TT cores and the accuracy of computed moments. To this end, we propose a quantitative measure that encapsulates these effects in the next section.


    \section{Contribution 3: Proposal of Score} \label{sec:score}
    As discussed in the previous section, the ordering of TT cores influences the accuracy of the computed moments. Based on the observations from the results in Sect.~\ref{sec:order_results}, we examine how the placement of interacting indices within the ordering of the TT cores affects accuracy, particularly when these indices are positioned on opposite sides of a partition that divides the TT cores into two groups. To systematically assess this effect, we introduce a new quantity called \textit{score} in this section.
    \subsection{Score}
    To quantify the impact of orderings of the TT cores on the accuracy of computed moments, we introduce a new quantity, referred to as  \textit{score}, for a given ordering of the TT cores
    $\left\{ \mathcal{T}^{(\sigma(i))}\right\}_{i=1}^d=\left\{ \mathcal{T}^{(\sigma(1))}, \mathcal{T}^{(\sigma(2))},\cdots,\mathcal{T}^{(\sigma(d))}\right\}$,
    where $\sigma$ is a permutation function that determines the ordering of the TT cores in the TT representation of the solution $\mathcal{P}(t)$ of Eq.~\eqref{eq:CME}.
    For example, the permutation $\sigma=\bigl( \begin{smallmatrix} 1 & 2 & 3 & 4\\3 & 2 & 1 & 4 \end{smallmatrix} \bigl)$ reorders the TT cores such that the solution is represented as
    $\mathcal{P}(t)=\sum_{r_1,r_2,r_3}\mathcal{T}^{(3)}_{1,n_3,r_1}\mathcal{T}^{(2)}_{r_1,n_2,r_2}\mathcal{T}^{(1)}_{r_2,n_1,r_3}\mathcal{T}^{(4)}_{r_3,n_4}.$ 
    The score quantifies how the accuracy of the computed moments are disrupted when the TT cores are reordered in this manner.

    For a given ordering of the TT cores with the permutation $\sigma$ and the given two-body interaction parameters $[\mu_{ij}]$, the score is defined as
    \begin{align}
        \mathrm{s}_{\sigma} = \sum_{1\leq i, j \leq d} |\mu_{ij}|\ \delta(i,j)
    \end{align}
     where $\delta(i,j)$ is a function that determines whether an interaction crosses the border $m$, which separates the TT cores into two halves. This function is given by
    \begin{align}
        \delta (i,j) = 
        \begin{cases}
            1 & \textrm{if $\sigma^{-1}(i)\leq m$ and $\sigma^{-1}(j)> m$, vice versa,}
            \\
            0 & \mathrm{otherwise.}
        \end{cases}
    \end{align}
    The border $m$ is defined as follows:
    \begin{align}
         m=
        \begin{cases}
            d/2 & \textrm{if $d$ is even,}\\
            \lfloor d/2\rfloor, \ \lceil d/2\rceil 
            & \textrm{if $d$ is odd.}
        \end{cases}
    \end{align}
    For odd $d$, the TT cores cannot be split into two equal halves, so the border falls between two adjacent cores. In this case, we compute two scores using $m = \lfloor d/2 \rfloor$ and $m = \lceil d/2 \rceil$, and take their mean as the final score.

\begin{figure}[bpt]
        \centering
        \begin{tikzpicture}
            \foreach \i [count=\j from 1] in {3,2,1,4} {
                \node[circle, draw=green!50!black, thick, minimum size=1cm] 
                (U\i) at ({1.5*\j}, 0) {$\mathcal{T}^{(\i)}$};
            }
    
            \foreach \i/\j in {3/2,2/1,1/4} {
                \draw[blue, thick] (U\i) -- (U\j);
            }
            
            \draw[red, thick, densely dotted] (U2) to[bend left=40] node[above right] {\small $\mu_{12},\ \mu_{21}$}  (U1);
            \draw[black, thick, densely dashed] (U3) to[bend left=40] (U2);
            \draw[red, thick, densely dotted] (U3) to[bend left=50] 
            node[above right] {\small $\mu_{34},\ \mu_{43}$} (U4);

            
            \draw[thick] (3.75,1.55) -- (3.75,-0.8);
            
            \foreach \i in {3,2,1,4} {
                \draw[black] (U\i) --++ (0,-0.7);
            }
        \end{tikzpicture}
        \caption{(Color online) TT cores (nodes in the figure) ordered by a permutation $\sigma=\bigl(
            \begin{smallmatrix}
                1 & 2 & 3 & 4\\3 & 2 & 1 & 4
            \end{smallmatrix}
            \bigl)$. The dotted and dashed curves connecting two TT cores indicate the two-body interactions with strengths $\mu_{ij}\ (1\leq i,j \leq 4)$. Especially, the dotted curves show interactions that cross the border and contribute to the score, while the dashed one does not contribute to the score; the score for this ordering is $|\mu_{12}|+|\mu_{21}|+|\mu_{34}|+|\mu_{43}|$.}
        \label{fig:permuted_TT}
    \end{figure}
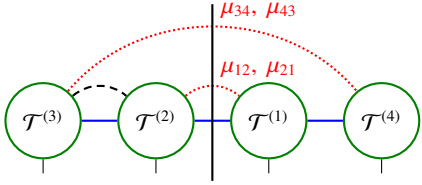

    For example, when $d=4$, the score is computed as the sum of $|\mu_{ij}|$ for all interactions between species $i$ and $j$ placed across the first two TT cores and the last two TT cores. If only nearest-neighbor interactions as in Eq.~\eqref{eq:params_dim4_cascade} are considered, the score for the ordering $\{\mathcal{T}^{(1)}, \mathcal{T}^{(2)}, \mathcal{T}^{(3)}, \mathcal{T}^{(4)}\}$ will be $\mathrm{s}_{\mathrm{id}}=|\mu_{23}|+|\mu_{32}|$, and that for the ordering $\{\mathcal{T}^{(3)}, \mathcal{T}^{(2)}, \mathcal{T}^{(1)}, \mathcal{T}^{(4)}\}$ with $\sigma=\bigl(
            \begin{smallmatrix}
                1 & 2 & 3 & 4\\3 & 2 & 1 & 4
            \end{smallmatrix}
            \bigl)$ will be $\mathrm{s}_{\sigma}=|\mu_{12}|+|\mu_{21}|+|\mu_{34}|+|\mu_{43}|$ as shown in Fig.~\ref{fig:permuted_TT}.

    \subsection{Experimental settings}
    We conducted experiments for $d = 4, 5$, and $6$, starting with the cascade process using the parameter set in Eq.~\eqref{eq:params_dim4_cascade} with no diffusion. Additionally, we performed simulations with non-uniform two-body interaction strengths as specified in Eq.~\eqref{eq:params_dim5_rand} with $[\bm{\sigma}]_i=0.25\ (1\leq i \leq d)$. For each case, we computed all first and second moments for $d!/2$ orderings of the TT cores, leveraging the reflection symmetry discussed in the previous section. The initial values were set as $\bm x_0=1.1 \cdot \bm 1$. 
    For the model without diffusion, reference values were computed via numerical integration with the Euler method, using a time step of $\Delta t = 1.0 \times 10^{-3}$. When diffusion is included, we used the Euler-Maruyama method with $\Delta t = 1.0 \times 10^{-3}$, and reference values were obtained as the mean of 5000 Monte Carlo samples, where we confirmed that the values had converged at this sample size.
    As already mentioned in Sect.~\ref{sec:intro}, the conventional CN method cannot afford the higher dimensional system such as $d=5$.

    The coupled equations were solved using the ALS algorithm with truncated SVD in the sweep procedure, as described in Sect.~\ref{sec:time_evolution}. ALS was implemented with a maximum TT rank of $R_i = 5$ for all $i = 1,2,\dots,d$, and the maximum number of state-space variables was set to $N = 10$. Under these conditions, each middle TT core had dimensions $5 \times 20 \times 5$, while the TT cores at both ends had dimensions $1 \times 20 \times 5$.

    
    \subsection{Numerical results}
\begin{figure}
        \centering
        \includegraphics[width=0.7\columnwidth]{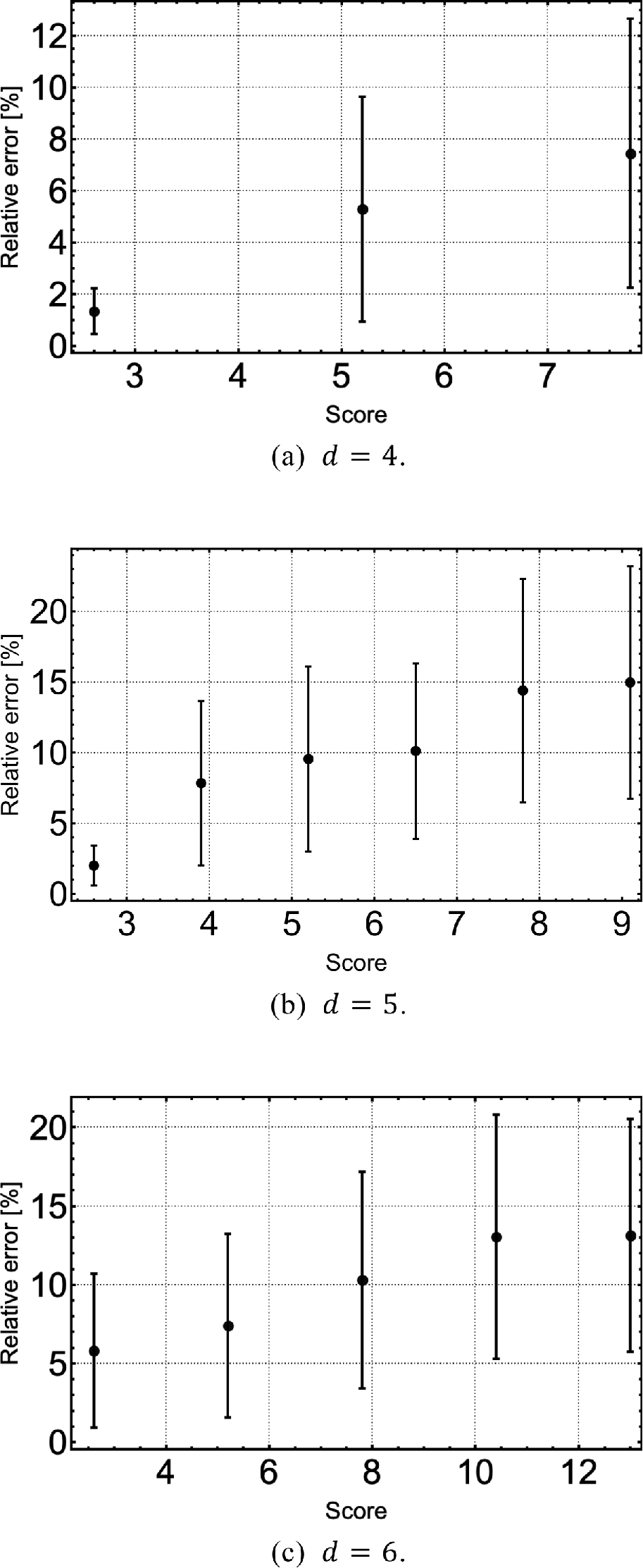}
        \caption{Relative errors for scores computed from orderings of the TT cores and the parameter set in Eq.~\eqref{eq:params_dim4_cascade} for $d = 4, 5$, and $6$ ((a), (b), and (c), respectively). The relative errors are computed for all the first and second moments. For each moment, we consider $d!/2$ distinct orderings due to reflected symmetry. Since the score is defined for a specific ordering of the TT cores, multiple relative errors correspond to each score. Moreover, the same score can be assigned to multiple orderings. The markers represent the mean relative error, with error bars indicating the standard deviation across these cases.}
        \label{fig:rel_err_cascades45}
    \end{figure}

\begin{figure}[ht]
        \centering
        \includegraphics[width=0.7\linewidth]{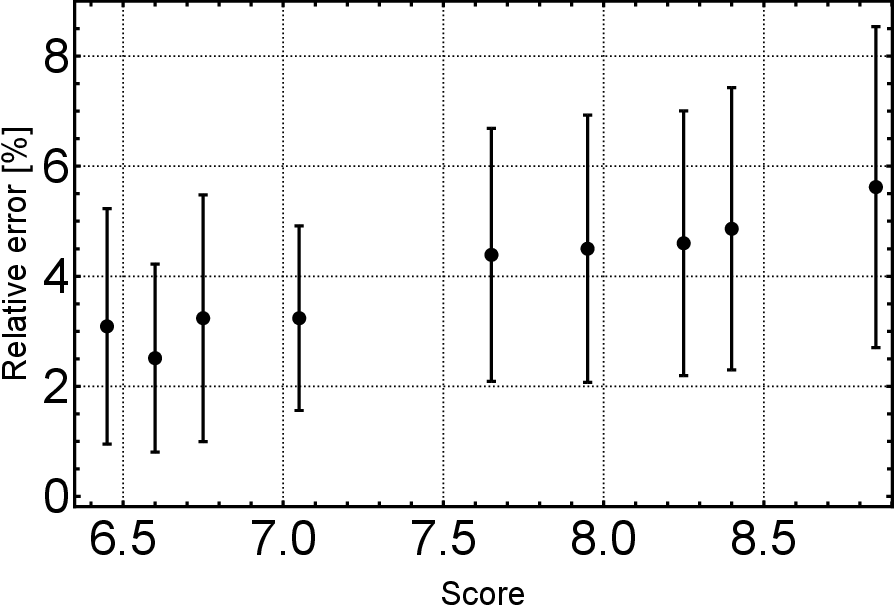}
        \caption{Relative errors plotted for scores, which are defined for orderings of the TT cores and the parameters set in Eq.~\eqref{eq:params_dim5_rand}. The relative errors are computed for all the first and second moments. Due to the reflected symmetry, the number of orderings of the TT cores considered is $d!/2$ distinct permutations. Since the score is defined for an ordering of the TT cores, multiple relative errors correspond to each score. Moreover, the same score can be assigned to multiple orderings. The markers and error bars have the same meanings as in Fig.~\ref{fig:rel_err_cascades45}}
        \label{fig:0ebefdd_score}
    \end{figure}

    Figures~\ref{fig:rel_err_cascades45} and~\ref{fig:0ebefdd_score} illustrate the relationships between the proposed score and the relative error for different settings.
    For each case, we computed the relative error for each ordering of the TT cores against the reference computed via the numerical integration or the MC sampling. The relative errors are plotted against the corresponding score. The results consistently show a positive correlation between the score and the relative error, with higher scores generally leading to larger errors. This trend confirms that the proposed score effectively reflects the accuracy of the computed moments across different orderings of the TT cores.

    Notably, for $d=4$ (Fig.~\ref{fig:rel_err_cascades45} (a)), the relative errors remain below 10\% across all tested orderings, with a clear increasing trend as the score increases. A similar pattern is observed for $d=5$ (Fig.~\ref{fig:rel_err_cascades45} (b)), though with a slightly larger spread of errors. In the case of $d=6$ (Fig.~\ref{fig:rel_err_cascades45} (c)), the error distribution is more dispersed, suggesting that the impact of ordering of the TT cores becomes more pronounced as $d$ increases. Finally, when considering randomly chosen interaction strengths (Fig.~\ref{fig:0ebefdd_score}), the relative errors are still influenced by the score, though the correlation appears slightly weaker compared to the cascade process.

    These findings suggest that the proposed score serves as a meaningful predictor of numerical accuracy, particularly for structured interactions as in the cascade process. However, in cases with more complex, randomly chosen interactions, additional factors may contribute to the observed error variations.



    \section{Conclusion}
    In this work, we first proposed the usage of the TT formats to evaluate the moments of the SDEs, which enables us to avoid the Monte Carlo samplings even in higher dimensional cases. Then, we examined the effect of ordering of the TT cores on the accuracy of moment computations in the SDEs. Using the duality relation in stochastic processes, we demonstrated that TT core arrangements significantly influence numerical accuracy, with certain orderings yielding systematically smaller relative errors. Our numerical experiments on the noisy LV model revealed that orderings of the TT cores preserving adjacent interactions tend to enhance accuracy, while non-adjacent configurations lead to large errors.

    To quantify these effects, we introduced a novel measure, the score, which reflects the degree to which interacting indices are separated in the TT representation. Our results indicate that minimizing this score generally leads to improved numerical accuracy. Furthermore, the proposed score serves as a practical heuristic for guiding ordering of the TT cores decisions, particularly in cases where interaction structures are complex.

    While this study provides a foundational understanding of effects of ordering the TT cores, future research should explore the permutations of the TT cores under the algorithm with rank adaptation.
    Moreover, given that coupled differential equations have been solved using the quantics TT (QTT) rather than the TT format \cite{dolgov2015simultaneous,kazeev2014direct}, investigating orderings in the QTT representation could offer further insights. 
    Additionally, extending these findings to higher-dimensional systems and other tensor network formats could further enhance the applicability of TT-based numerical methods in high-dimensional stochastic systems.

\begin{acknowledgments}
R. S. is supported by the JSPS KAKENHI Grant No. 23KJ0295. This work was supported by JST FOREST Program (JST Grant Number JPMJFR216K) and by the Center of Innovation for Sustainable Quantum AI (JST Grant Number JPMJPF2221).
\end{acknowledgments}





\bibliographystyle{apsrev4-2}
\bibliography{ref}

\end{document}